\documentclass[letterpaper,preprintnumbers,twocolumn,eqsecnum,superscriptaddress,aps,nofootinbib]{revtex4}
\usepackage{amssymb}
\usepackage[centertags]{amsmath}
\usepackage{txfonts}
\usepackage{epsfig}
\usepackage{bm}
\usepackage{color}
\usepackage{graphicx,graphics}
\usepackage{multirow}
\usepackage{float}
\usepackage[pdfstartview=FitH]{hyperref}
\hypersetup{colorlinks=true, citecolor=blue, linkcolor=blue,filecolor=black,urlcolor=blue}
\allowdisplaybreaks[2]
\usepackage{setspace}
\usepackage{slashed}
\usepackage{ulem}

\usepackage{booktabs}
\usepackage{multirow}

\begin{document}

\title{Twist-4 contributions to semi-inclusive $e^+e^-$-annihilation process}

\author{Wei-hua Yang}
\affiliation{School of Physics \&  Key Laboratory of Particle Physics and Particle Irradiation (MOE),
  Shandong University, Jinan, Shandong 250100, China }

\author{Kai-bao Chen}
\affiliation{School of Physics \&  Key Laboratory of Particle Physics and Particle Irradiation (MOE),
  Shandong University, Jinan, Shandong 250100, China }

\author{Zuo-tang Liang}
\affiliation{School of Physics \&  Key Laboratory of Particle Physics and Particle Irradiation (MOE),
  Shandong University,  Jinan, Shandong 250100, China }

\begin{abstract}
We present the complete twist-4 results for the semi-inclusive annihilation process $e^+ +e^- \to h+ \bar q+X$ at the tree level
of perturbative quantum chromodynamics.
The calculations are carried out by using the formalism obtained by applying the collinear expansion to this process
where the multiple gluon scattering is taken into account and gauge links are obtained systematically and automatically.
We present the results for structure functions in terms of gauge invariant fragmentation functions up to twist-4 and the corresponding results
for the azimuthal asymmetries and polarizations of hadrons produced.
The results obtained show in particular that similar to that for semi-inclusive deeply inelastic scattering,
for structure functions associated with
sine or cosine of odd number of azimuthal angle(s), there are only twist-3 contributions,
while for those of even number of azimuthal angle(s),
there are leading twist and twist-4 contributions.
For all those structure functions that have leading twist contributions, there are twist-4 addenda to them.
Hence twist-4 contributions may even have large influences on extracting leading twist fragmentation functions from the data.
We also suggest a method for a rough estimation of twist-4 contributions based on the leading twist fragmentation functions.
\end{abstract}


\maketitle

\section{Introduction}

Parton distribution functions (PDFs) and fragmentation functions (FFs) are two important quantities in describing high-energy reactions.
When three-dimensional, i.e., the transverse momentum dependent (TMD) PDFs and/or FFs are considered,
the sensitive quantities studied in experiments are
often different azimuthal asymmetries~\cite{Airapetian:1999tv,Airapetian:2004tw,Airapetian:2009ae,Airapetian:2010ds,Airapetian:2012yg, Alexakhin:2005iw,Ageev:2006da,Alekseev:2008aa,Alekseev:2010rw,Adolph:2012sn,Adolph:2012sp,Adolph:2014pwc,Adolph:2014zba,Avakian:2010ae, Aghasyan:2011ha,Qian:2011py,Huang:2011bc,Zhang:2013dow,Zhao:2014qvx,
Abe:2005zx,Seidl:2008xc,Vossen:2011fk,TheBABAR:2013yha,Ablikim:2015pta,Barone:2010zz,Liang:2015nia}. 
In such cases higher twist contributions can be very significant and play a very important role in studying these TMD PDFs and/or FFs.
Twist-3 contributions to semi-inclusive deeply inelastic scattering (SIDIS)~\cite{Mulders:1995dh,Boer:1997nt,Bacchetta:2004zf,Bacchetta:2006tn,Liang:2006wp,Song:2010pf,Song:2013sja,Yang:2016qsf}
and $e^+e^-$-annihilations~\cite{Boer:1997mf,Boer:2008fr,Wei:2013csa,Wei:2014pma,Chen:2016moq,Pitonyak:2013dsu}
have been extensively calculated in recent years.
Results have been given for cross section and different azimuthal asymmetries in terms of gauge invariant PDFs and/or FFs.

In a recent paper, we have carried out the calculations of twist-4 contributions to the SIDIS process $e^-N\to e^-qX$ \cite{Wei:2016far}.
The results obtained show a very distinct feature, i.e.,
while all twist-3 contributions lead to azimuthal asymmetries absent at leading twist,
the twist-4 contributions are just addenda to the leading twist asymmetries.
We have twist-4 contributions to all the eight leading twist structure functions corresponding to the eight leading twist TMD PDFs.
This implies that studying twist-4 contributions is important not only for itself but also in determining leading twist TMD PDFs.
It may leads to significant modifications in extracting leading twist PDFs from experimental data~\cite{Airapetian:1999tv,Airapetian:2004tw,Airapetian:2009ae,Airapetian:2010ds,Airapetian:2012yg, Alexakhin:2005iw,Ageev:2006da,Alekseev:2008aa,Alekseev:2010rw,Adolph:2012sn,Adolph:2012sp,Adolph:2014pwc,Adolph:2014zba,Avakian:2010ae, Aghasyan:2011ha,Qian:2011py,Huang:2011bc,Zhang:2013dow,Zhao:2014qvx}.

While three-dimensional PDFs are best studied in SIDIS, three-dimensional FFs are best studied in the semi-inclusive process $e^+e^-\to h\bar qX$.
Moreover, we can study not only the vector polarization dependent FFs but also tensor polarization dependent FFs.
In view of the conclusions presented in~\cite{Wei:2014pma,Chen:2016moq}, it is natural and important to extend the twist-4 calculations to $e^+e^-\to h\bar qX$.

In this paper, we present the twist-4 studies to the semi-inclusive annihilation process $e^+e^-\to h\bar qX$.
We present the complete calculations at tree level in perturbative  quantum chromodynamics (pQCD) and show the results
of the structure functions in terms of gauge invariant FFs.
We present the results for the unpolarized, the vector polarization dependent and tensor polarization dependent parts, respectively.
We also present the azimuthal asymmetries and hadron polarizations in terms of the gauge invariant FFs.

The higher twist calculations presented in, e.g.,~\cite{Liang:2006wp,Song:2013sja,Song:2010pf,Wei:2013csa,Wei:2014pma,Chen:2016moq,Wei:2016far}
benefited very much from the collinear expansion.
We found out that in dealing with higher twist effects in  quantum chromodynamics (QCD) parton model for high energy reactions,
collinear expansion is indeed extremely important and powerful.
It provides not only the correct formalism where the differential cross section or the hadronic tensor is given in terms of gauge invariant PDFs and/or FFs
but also very simplified expressions so that even twist-4 contributions can be calculated.
The collinear expansion has been first introduced in 1980 to 90s and has been applied to inclusive processes~\cite{Ellis:1982wd,Ellis:1982cd,Qiu:1990xxa,Qiu:1990xy}.
It has been shown that it can also be applied to the semi-inclusive DIS process $e^-N\to e^-qX$~\cite{Liang:2006wp},
and recently also to inclusive $e^+e^-\to hX$~\cite{Wei:2013csa} and  semi-inclusive  process $e^+e^-\to h\bar q X$~\cite{Wei:2014pma}.
As has been emphasized in~\cite{Liang:2006wp,Song:2013sja,Song:2010pf,Wei:2013csa,Wei:2014pma,Chen:2016moq,Wei:2016far},
the collinear expansion is a necessary procedure for obtaining hadronic tensor in terms of gauge-invariant PDFs and/or FFs.
Moreover, the hard parts after the collinear expansion are not only calculable but also reduced to a form independent of the parton momenta besides some delta-function.
Correspondingly, the involved PDFs and/or FFs are not only gauge invariant but also all defined via quark-quark or quark-$j$-gluon-quark correlator with one independent parton momentum.
Hence the Lorentz decomposition of such quark-quark or quark-$j$-gluon-quark correlator is feasible and higher twist calculations can be carried out.

The rest of the paper is organized as follows.
In Sec.~\ref{sec:formalism}, we present the formalism of $e^+ e^- \rightarrow h \bar qX$
where we show both the results of the general kinematic analysis and those for the hadronic tensor in QCD parton model after collinear expansion.
In Sec.~\ref{sec:results}, we present the results for the hadronic tensor, the structure functions, 
the azimuthal asymmetries and hadron polarizations at the tree level up to twist-4.
A short summary and discussion is given in Sec.~\ref{sec:summary}.

\section{The formalism}\label{sec:formalism}

\begin{figure}
  \includegraphics[width=6.0cm]{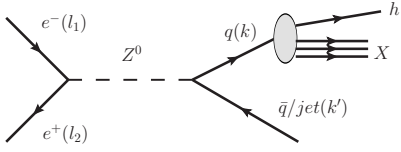}\\
  \caption{Illustrating diagram for $e^++e^-\to h+\bar q+X$.}\label{fig:ff1}
\end{figure}

To be explicit, we consider the semi-inclusive process $e^+ e^- \to Z^0\to h \bar qX$
where $\bar q$ denotes an anti-quark that corresponds to a jet of hadrons in experiments and $h$ denotes the outgoing hadron.
The cross section is given by
\begin{align}
\frac{2E_pE_{k'} d\sigma}{d^3p d^3k'} = \frac{\alpha^2}{8\pi^3sQ^4} \chi L^{\mu\nu}(l_1,l_2) W_{\mu\nu}^{(si)}(q,p,S,k').
\end{align}
Here we use the notations as illustrated in Fig.~\ref{fig:ff1};
$\alpha = e^2/4\pi$, $\chi = Q^4/[(Q^2-M_Z^2)^2 + \Gamma_Z^2 M_Z^2]\sin^4 2\theta_W$, $Q^2 = s = q^2$, $\theta_W$ is the Weinberg angle,
$M_Z$ is the $Z$-boson mass and $\Gamma_Z$ is the decay width.
The leptonic tensor is given by
\begin{align}
L_{\mu\nu}(l_1,l_2) =& c_1^e(l_{1\mu} l_{2\nu}+l_{1\nu} l_{2\mu}-g_{\mu\nu}l_1\cdot l_2) + ic_3^e\varepsilon_{\mu\nu l_{1}l_{2}},
\label{eq:leptonictensor}
\end{align}
where $c_1^e = (c_V^e)^2 + (c_A^e)^2$ and $c_3^e = 2 c_V^e c_A^e$;
$c_V^e$ and $c_A^e$ are defined in the weak interaction current
$J_\mu (x)=\bar \psi(x)\Gamma_\mu\psi(x)$ and $\Gamma_\mu= \gamma_\mu (c_V^e - c_A^e \gamma^5)$.
Similar notations are also used for quarks where we use a superscript $q$ to replace $e$.
We use also the shorthand notations such as $\varepsilon_{\mu\nu AB}\equiv \varepsilon_{\mu\nu\alpha\beta} A^\alpha B^\beta$.
The hadronic tensor is defined as
\begin{align}
W_{\mu\nu}^{(si)} &(q,p,S,k') = \frac{1}{2\pi}  \sum_X (2\pi)^4 \delta^4 (q-p-k'- p_X) \nonumber\\
& \times \langle 0| J_\nu (0) |p,S,k';X\rangle \langle p,S,k';X |J_\mu (0)|0\rangle,
\end{align}
where $S$ denotes the polarization of the hadron and $J_\mu (x)$ is the quark electroweak current.
It is related to the inclusive hadronic tensor $W_{\mu\nu}^{(in)}(q,p,S)$ by
\begin{align}
W_{\mu\nu}^{(in)} &(q,p,S) = \int \frac{d^3k'}{(2\pi)^3 2E_{k'}} W_{\mu\nu}^{(si)}(q,p,S,k').
\end{align}
If we consider only the transverse momentum $k_\perp'$ dependence, we have
\begin{align}
\frac{E_pd\sigma}{d^3p d^2k'_\perp} = \frac{\alpha^2 \chi}{4\pi^2sQ^4} L^{\mu\nu}(l_1,l_2) W_{\mu\nu}(q,p,S,k'_\perp),
\end{align}
where $W_{\mu\nu}(q,p,S,k'_\perp)$ is the TMD semi-inclusive hadronic tensor given by
\begin{align}
W_{\mu\nu} &(q,p,S,k'_\perp) = \int \frac{dk'_z}{(2\pi)2E_{k'}}  W_{\mu\nu}^{(si)}(q,p,S,k').
\end{align}

\subsection{The general form of the cross section in terms of structure functions}

Formally, the general form of the cross section for $e^+e^- \to h \bar qX$
is exactly the same as that for $e^+e^- \to V\pi X$ discussed in detail in~\cite{Chen:2016moq}.
We summarize the results here.

The hadronic tensor is divided into a symmetric and an anti-symmetric part,
$W_{\mu\nu} =W^S_{\mu\nu} + iW^A_{\mu\nu}$,
each of them is given by a linear combination of a set of basic Lorentz tensors (BLTs), i.e.,
\begin{align}
   W^{S\mu\nu} &=\sum_{\sigma,j} W_{\sigma j}^S h_{\sigma j}^{S\mu\nu} + \sum_{\sigma, j} \tilde W_{\sigma j}^S \tilde h_{\sigma j}^{S\mu\nu},\label{eq:Wsmunu}\\
   W^{A\mu\nu} &=\sum_{\sigma,j} W_{\sigma j}^A h_{\sigma j}^{A\mu\nu} + \sum_{\sigma, j} \tilde W_{\sigma j}^A \tilde h_{\sigma j}^{A\mu\nu},\label{eq:Wamunu}
\end{align}
where $h^{\mu\nu}$ and $\tilde h^{\mu\nu}$ represent the space reflection even and space reflection odd BLTs, respectively.
The subscript $\sigma$ specifies the polarization.

As has been found out in~\cite{Chen:2016moq}, a distinct feature for BLTs in semi-inclusive reactions such as $e^+e^- \to h \bar qX$
is that the polarization dependent BLTs can be taken as a product of the unpolarized BLTs and polarization dependent Lorentz scalar(s) or pseudo-scalar(s).
There are 9 unpolarized BLTs given by
\begin{align}
   h^{S\mu\nu}_{Ui}&=\Big\{g^{\mu\nu}-\frac{q^\mu q^\nu}{q^2}, ~p_q^\mu  p_q^\nu,~ k_q^{\prime\mu}  k_q^{\prime\nu},  ~ p_q^{\{\mu} k_q^{\prime\nu\}}\Big\},\label{eq:hsU} \\
   \tilde h^{S\mu\nu}_{Ui}&=\Big\{\varepsilon^{\mu q p k'} p_q^\nu,~ \varepsilon^{\mu q p k'}k_q^{\prime\nu}\Big\}, \label{eq:thsU} \\
   h^{A\mu\nu}_{U}&=\Big\{p_q^{[\mu} k_q^{\prime\nu]}\Big\},\label{eq:haU} \\
   \tilde h^{A\mu\nu}_{Ui}&=\Big\{\varepsilon ^{\mu\nu qp},~ \varepsilon ^{\mu\nu qk'}\Big\}.\label{eq:thaU}
\end{align}
The subscript $U$ denotes the unpolarized part.
Here $p_q\equiv p- q(p\cdot q)/q^2$ satisfying $p_q\cdot q=0$,
$A^{\{\mu}B^{\nu\}} \equiv A^\mu B^\nu +A^\nu B^\mu$,  and $A^{[\mu}B^{\nu]} \equiv A^\mu B^\nu -A^\nu B^\mu$.

The vector polarization dependent BLTs are given by
\begin{align}
  h^{S\mu\nu}_{Vi} &=\Big\{[\lambda_h, (k'_\perp\cdot S_T)]\tilde h^{S\mu\nu}_{Ui}, ~\varepsilon_\perp^{k'S} h^{S\mu\nu}_{Uj}\Big\}, \label{eq:hsV}\\
 \tilde h^{S\mu\nu}_{Vi} &=\Big\{[\lambda_h, (k'_\perp\cdot S_T)]h^{S\mu\nu}_{Ui}, ~ \varepsilon_\perp^{k'S} \tilde h^{S\mu\nu}_{Uj} \Big\}, \label{eq:thsV}\\
  h^{A\mu\nu}_{Vi} &=\Big\{[\lambda_h, (k'_\perp\cdot S_T)]\tilde h^{A\mu\nu}_{Ui},~ \varepsilon_\perp^{k'S}  h^{A\mu\nu}_{U}\Big\},\label{eq:haV}\\
  \tilde h^{A\mu\nu}_{Vi} &=\Big\{[\lambda_h, (k'_\perp\cdot S_T)]h^{A\mu\nu}_{U},~ \varepsilon_\perp^{k'S}  \tilde h^{A\mu\nu}_{Uj}\Big\},\label{eq:thaV}
\end{align}
where $\varepsilon_\perp^{k'S}=\varepsilon_\perp^{\alpha\beta} k'_\alpha S_\beta$,
$\varepsilon_\perp^{\alpha\beta}=\varepsilon^{\mu\nu\alpha\beta}\bar n_\mu n_\nu$;
$\lambda_h$ is the helicity of the outgoing hadron and $S_T$ denotes the transverse polarization components.
There are 27 such vector polarized BLTs in total.

The tensor polarized part is composed of $S_{LL}$-, $S_{LT}$- and $S_{TT}$-dependent parts.
There are 9 $S_{LL}$-dependent BLTs, they are given by
\begin{align}
&h_{LLi}^{S\mu\nu}=S_{LL} h^{S\mu\nu}_{Ui}, &&\tilde h_{LLi}^{S\mu\nu}=S_{LL} \tilde h^{S\mu\nu}_{Ui}, \label{eq:hsLL} \\
&h_{LL}^{A\mu\nu}=S_{LL} h^{A\mu\nu}_{U}, &&\tilde h_{LLi}^{A\mu\nu}=S_{LL} \tilde h^{A\mu\nu}_{Ui}. \label{eq:thsLL}
\end{align}
There are 18 $S_{LT}$-dependent ones, they are
\begin{align}
 & h^{S\mu\nu}_{LTi} =\Big\{ (k'_\perp\cdot S_{LT}) h^{S\mu\nu}_{Ui}, ~\varepsilon_\perp^{k'S_{LT}} \tilde h^{S\mu\nu}_{Uj}\Big\}, \label{eq:hsLT}\\
 & \tilde h^{S\mu\nu}_{LTi} =\Big\{(k'_\perp\cdot S_{LT})\tilde h^{S\mu\nu}_{Ui}, ~ \varepsilon_\perp^{k'S_{LT}} h^{S\mu\nu}_{Uj} \Big\}, \label{eq:thsLT}\\
 & h^{A\mu\nu}_{LTi} =\Big\{(k'_\perp\cdot S_{LT}) h^{A\mu\nu}_{U},~ \varepsilon_\perp^{k'S_{LT}}  \tilde h^{A\mu\nu}_{Uj}\Big\},\label{eq:haLT}\\
 & \tilde h^{A\mu\nu}_{LTi} =\Big\{(k'_\perp\cdot S_{LT})\tilde h^{A\mu\nu}_{Ui},~ \varepsilon_\perp^{k'S_{LT}} h^{A\mu\nu}_{U}\Big\}. \label{eq:thaLT}
\end{align}
There are also 18 $S_{TT}$-dependent ones,  they are
\begin{align}
 & h^{S\mu\nu}_{TTi} =\Big\{ S_{TT}^{k'k'} h^{S\mu\nu}_{Ui}, ~S_{TT}^{\tilde k'k'}  \tilde h^{S\mu\nu}_{Uj}\Big\}, \label{eq:hsTT}\\
 & \tilde h^{S\mu\nu}_{TTi} =\Big\{S_{TT}^{k'k'} \tilde h^{S\mu\nu}_{Ui}, ~ S_{TT}^{\tilde k'k'} h^{S\mu\nu}_{Uj} \Big\}, \label{eq:thsTT}\\
 & h^{A\mu\nu}_{TTi} =\Big\{S_{TT}^{k'k'}  h^{A\mu\nu}_{U},~ S_{TT}^{\tilde k'k'}  \tilde h^{A\mu\nu}_{Uj}\Big\}, \label{eq:haTT}\\
 & \tilde h^{A,\mu\nu}_{TTi} =\Big\{S_{TT}^{k'k'} \tilde h^{A\mu\nu}_{Ui},~ S_{TT}^{\tilde k'k'} h^{A\mu\nu}_{U}\Big\},\label{eq:thaTT}
\end{align}
where $S_{TT}^{k'k'}=k'_{\perp\alpha} S_{TT}^{\alpha\beta} k'_{\perp\beta}$ and $\tilde k'_{\perp\alpha}=\varepsilon_{\perp \alpha k'}$.
There are  in total 81 such BLTs. Correspondingly there should be 81 structure functions for $e^+e^-\to h\bar qX$.

The cross section is given in the helicity Gottfried-Jackson frame~\cite{Chen:2016moq} where we choose
$p=(E_p,0,0,p_z)$, and $l_1=Q(1,\sin\theta,0,\cos\theta)/2$,
$k'=(E_{k'}, |\vec k'_\perp|\cos\varphi,|\vec k'_\perp|\sin\varphi,k'_z)$ and
\begin{align}
&S=(\lambda_h\frac{p_z}{M},|S_T|\cos\varphi_S, |S_T|\sin\varphi_S,\lambda_h\frac{E_P}{M}), \\
&S_{LT}=(0,|S_{LT}|\cos\varphi_{LT}, |S_{LT}|\sin\varphi_{LT},0), \\
&S_{TT}^{x\mu}= (0,|S_{TT}|\cos2\varphi_{TT}, |S_{TT}|\sin2\varphi_{TT},0).
\end{align}
In this frame, the cross section is given by
\begin{align}
 & \frac{E_p d\sigma}{d^3 p d^2k'_\perp}=\frac{\alpha^2\chi}{4\pi^2s^2}\Bigl[
 (\mathcal{W}_U+ \tilde{\mathcal{W}}_U) +\lambda_h (\mathcal{W}_L+ \tilde{\mathcal{W}}_L)\nonumber\\
&~~~~~~+|S_T|(\mathcal{W}_T+ \tilde{\mathcal{W}}_T)
+S_{LL}(\mathcal{W}_{LL}+ \tilde{\mathcal{W}}_{LL})\nonumber\\
&~~~~~~+|S_{LT}|(\mathcal{W}_{LT}+ \tilde{\mathcal{W}}_{LT})
+|S_{TT}|(\mathcal{W}_{TT}+ \tilde{\mathcal{W}}_{TT})\Bigr], \label{eq:cs}
\end{align}
where we use $\mathcal{W}_\sigma$ and $\tilde {\mathcal{W}}_\sigma$ to denote the parity conserved and parity violated parts, respectively.
For the unpolarized part, they are given by
\begin{align}
 \mathcal{W}_U =&(1+\cos^2\theta)W_{U1}+\sin^2\theta W_{U2}+ \cos\theta W_{U3}\nonumber\\
  &+\cos\varphi [\sin\theta W^{\cos\varphi}_{U1}+\sin2\theta W^{\cos\varphi}_{U2}]\nonumber\\
  &+\cos2\varphi\sin^2\theta W^{\cos2\varphi}_U, \label{eq:Wu}\\
  \tilde{\mathcal{W}}_U =&\sin\varphi [\sin\theta\tilde W^{\sin\varphi}_{U1}+\sin2\theta\tilde W^{\sin\varphi}_{U2}]\nonumber\\
  &+\sin2\varphi\sin^2\theta\tilde W^{\sin2\varphi}_U. \label{eq:tWu}
\end{align}
We note in particular that all the $\theta$ and $\varphi$ dependences are given explicitly.
The $W_{Uj}$ and $\tilde W_{Uj}$ are scalar functions depending on
$s$, $\xi=2q\cdot p/q^2$ and $k_\perp^{\prime2}$ and are called ``structure functions''.
The subscript $j$ specifies different $\theta-$dependence modes for the same $\varphi-$dependence.
We have 6 structure functions corresponding to parity conserved terms and 3 corresponding to parity violated terms in the unpolarized case.

The longitudinal polarization (helicity $\lambda_h$ and spin alignment $S_{LL}$) dependent parts take exactly the same form as the unpolarized part.
More precisely, $\tilde{\mathcal{W}}_L$ and $\mathcal{W}_{LL}$ take the same form as $\mathcal{W}_U$; $\mathcal{W}_L$ and $\tilde{\mathcal{W}}_{LL}$ take the same form as $\tilde{\mathcal{W}}_U$,
 i.e.,
\begin{align}
  \mathcal{W}_L =&\sin\varphi [\sin\theta W^{\sin\varphi}_{L1}+\sin2\theta W^{\sin\varphi}_{L2} ]\nonumber\\
  &+\sin2\varphi\sin^2\theta  W^{\sin2\varphi}_L, \label{eq:WL}\\
  \tilde{\mathcal{W}}_L =&(1+\cos^2\theta)\tilde W_{L1}+\sin^2\theta \tilde W_{L2}+ \cos\theta \tilde W_{L3}\nonumber\\
  &+\cos\varphi\big[\sin\theta \tilde W^{\cos\varphi}_{L1}+\sin2\theta \tilde W^{\cos\varphi}_{L2}\big]\nonumber\\
  &+\cos2\varphi\sin^2\theta \tilde W^{\cos2\varphi}_L; \label{eq:tWL} \\
  \mathcal{W}_{LL} =&(1+\cos^2\theta)W_{LL1}+\sin^2\theta W_{LL2}+ \cos\theta W_{LL3} \nonumber\\
   &+\cos\varphi[\sin\theta W^{\cos\varphi}_{LL1}+\sin2\theta W^{\cos\varphi}_{LL2}] \nonumber\\
   &+\cos2\varphi\sin^2\theta W^{\cos2\varphi}_{LL}, \label{eq:WLL}\\
  \tilde{\mathcal{W}}_{LL} =&\sin\varphi [\sin\theta \tilde W^{\sin\varphi}_{LL1}+\sin2\theta \tilde W^{\sin\varphi}_{LL2} ] \nonumber\\
   &+\sin2\varphi\sin^2\theta \tilde W^{\sin2\varphi}_{LL}. \label{eq:tWLL}
\end{align}
For the transverse polarization dependent parts, there are three azimuthal angles, $\varphi_S$, $\varphi_{LT}$ and $\varphi_{TT}$  involved.
The corresponding expressions are
\begin{align}
  \mathcal{W}_T &= \sin\varphi_S [\sin\theta W^{\sin\varphi_S}_{T1}+\sin2\theta W^{\sin\varphi_S}_{T2}]\nonumber\\
  &+\sin(\varphi+\varphi_S)\sin^2\theta W^{\sin(\varphi+\varphi_S)}_T \nonumber\\
  &+\sin(\varphi-\varphi_S)[(1+\cos^2\theta)W^{\sin(\varphi-\varphi_S)}_{T1} \nonumber\\
      &\hspace{1cm}+\sin^2\theta W^{\sin(\varphi-\varphi_S)}_{T2}+ \cos\theta W^{\sin(\varphi-\varphi_S)}_{T3}]  \nonumber\\
  &+\sin(2\varphi-\varphi_S)[\sin\theta W^{\sin(2\varphi-\varphi_S)}_{T1}+\sin2\theta W^{\sin(2\varphi-\varphi_S)}_{T2}] \nonumber\\
  &+\sin(3\varphi-\varphi_S)\sin^2\theta W^{\sin(3\varphi-\varphi_S)}_T,  \label{eq:WT} \\
  \tilde{\mathcal{W}}_T &= \cos\varphi_S [\sin\theta \tilde W^{\cos\varphi_S}_{T1}+\sin2\theta \tilde W^{\cos\varphi_S}_{T2}] \nonumber\\
  &+\cos(\varphi+\varphi_S)\sin^2\theta \tilde W^{\cos(\varphi+\varphi_S)}_T \nonumber\\
  &+\cos(\varphi-\varphi_S)[(1+\cos^2\theta)\tilde W^{\cos(\varphi-\varphi_S)}_{T1} \nonumber\\
      &\hspace{1cm}+\sin^2\theta \tilde W^{\cos(\varphi-\varphi_S)}_{T2}+ \cos\theta \tilde W^{\cos(\varphi-\varphi_S)}_{T3}]  \nonumber\\
  &+\cos(2\varphi-\varphi_S) [\sin\theta \tilde W^{\cos(2\varphi-\varphi_S)}_{T1}+\sin2\theta \tilde W^{\cos(2\varphi-\varphi_S)}_{T2}] \nonumber\\
  &+\cos(3\varphi-\varphi_S)\sin^2\theta \tilde W^{\cos(3\varphi-\varphi_S)}_T;  \label{eq:tWT} \\
  \tilde{\mathcal{W}}_{LT} &= \sin\varphi_{LT}[\sin\theta \tilde W^{\sin\varphi_{LT}}_{LT1}+\sin2\theta \tilde W^{\sin\varphi_{LT}}_{LT2}] \nonumber\\
  &+\sin(\varphi+\varphi_{LT})\sin^2\theta \tilde W^{\sin(\varphi+\varphi_{LT})}_{LT} \nonumber\\
  &+\sin(\varphi-\varphi_{LT})[(1+\cos^2\theta)\tilde W^{\sin(\varphi-\varphi_{LT})}_{LT1}\nonumber\\
      &\hspace{1cm}+\sin^2\theta \tilde W^{\sin(\varphi-\varphi_{LT})}_{LT2}+ \cos\theta \tilde W^{\sin(\varphi-\varphi_{LT})}_{LT3}]  \nonumber\\
  &+\sin(2\varphi-\varphi_{LT})[\sin\theta \tilde W^{\sin(2\varphi-\varphi_{LT})}_{LT1}+\sin2\theta \tilde W^{\sin(2\varphi-\varphi_{LT})}_{LT2}] \nonumber\\
  &+\sin(3\varphi-\varphi_{LT})\sin^2\theta \tilde W^{\sin(3\varphi-\varphi_{LT})}_{LT},  \label{eq:WLT} \\
  \mathcal{W}_{LT}&= \cos\varphi_{LT}[\sin\theta W^{\cos\varphi_{LT}}_{LT1}+\sin2\theta  W^{\cos\varphi_{LT}}_{LT2}]\nonumber\\
  &+\cos(\varphi+\varphi_{LT})\sin^2\theta  W^{\cos(\varphi+\varphi_{LT})}_{LT} \nonumber\\
  &+\cos(\varphi-\varphi_{LT})[(1+\cos^2\theta) W^{\cos(\varphi-\varphi_{LT})}_{LT1}\nonumber\\
      &\hspace{1cm}+\sin^2\theta  W^{\cos(\varphi-\varphi_{LT})}_{LT2}+ \cos\theta  W^{\cos(\varphi-\varphi_{LT})}_{LT3}]  \nonumber\\
  &+\cos(2\varphi-\varphi_{LT})[\sin\theta  W^{\cos(2\varphi-\varphi_{LT})}_{LT1}+\sin2\theta W^{\cos(2\varphi-\varphi_{LT})}_{LT2}]\nonumber\\
  &+\cos(3\varphi-\varphi_{LT})\sin^2\theta W^{\cos(3\varphi-\varphi_{LT})}_{LT};  \label{eq:tWLT} \\
  \tilde{\mathcal{W}}_{TT} &=\sin(\varphi-2\varphi_{TT})[\sin\theta \tilde W^{\sin(\varphi-2\varphi_{TT})}_{TT1}+\sin2\theta \tilde W^{\sin(\varphi-2\varphi_{TT})}_{TT2}]\nonumber\\
  &+\sin2\varphi_{TT}\sin^2\theta \tilde W^{\sin2\varphi_{TT}}_{TT} \nonumber\\
  &+\sin(2\varphi-2\varphi_{TT})[(1+\cos^2\theta)\tilde W^{\sin(2\varphi-2\varphi_{TT})}_{TT1}\nonumber\\
      &\hspace{1cm}+\sin^2\theta \tilde W^{\sin(2\varphi-2\varphi_{TT})}_{TT2}+ \cos\theta \tilde W^{\sin(2\varphi-2\varphi_{TT})}_{TT3}]  \nonumber\\
  &+\sin(3\varphi-2\varphi_{TT})[\sin\theta \tilde W^{\sin(3\varphi-2\varphi_{TT})}_{TT1}+\sin2\theta \tilde W^{\sin(3\varphi-2\varphi_{TT})}_{TT2}]\nonumber\\
  &+\sin(4\varphi-2\varphi_{TT})\sin^2\theta \tilde W^{\sin(4\varphi-2\varphi_{TT})}_{TT},  \label{eq:WTT} \\
  \mathcal{W}_{TT}&= \cos(\varphi-2\varphi_{TT})[\cos\theta W^{\cos(\varphi-2\varphi_{TT})}_{TT1}+\sin2\theta  W^{\cos(\varphi-2\varphi_{TT})}_{TT2}]\nonumber\\
  &+\cos2\varphi_{TT}\sin^2\theta  W^{\cos2\varphi_{TT}}_{TT} \nonumber\\
  &+\cos(2\varphi-2\varphi_{TT})[(1+\cos^2\theta) W^{\cos(2\varphi-2\varphi_{TT})}_{TT1}\nonumber\\
      &\hspace{1cm}+\sin^2\theta  W^{\cos(2\varphi-2\varphi_{TT})}_{TT2}+ \cos\theta  W^{\cos(2\varphi-2\varphi_{TT})}_{TT3}]  \nonumber\\
  &+\cos(3\varphi-2\varphi_{TT})[\sin\theta W^{\cos(3\varphi-2\varphi_{TT})}_{TT1}+\sin2\theta  W^{\cos(3\varphi-2\varphi_{TT})}_{TT2}]\nonumber\\
  &+\cos(4\varphi-2\varphi_{TT})\sin^2\theta  W^{\cos(4\varphi-2\varphi_{TT})}_{TT}.  \label{eq:tWTT}
\end{align}
We note in particular the following:
Since $S$ is an axial vector and $S_{LT}$ is a vector, we have one to one correspondence
between ${\mathcal W}_T\leftrightarrow \tilde{{\mathcal W}}_{LT}$ and $\tilde{\mathcal W}_T\leftrightarrow {{\mathcal W}}_{LT}$.
Also $S_{LT}^\mu$ corresponds to $S_{TT}^{k'\mu}$ in the BLTs given by Eqs.~(\ref{eq:hsLT})-(\ref{eq:thaTT}),
hence $\varphi_{LT}$ corresponds to $2\varphi_{TT}-\varphi$.

Having the general form of the differential cross section, we can express all other measurable quantities such as
azimuthal asymmetries and different components of hadron polarizations in terms of structure functions.
The longitudinal components are unique and defined with respect to the direction of the hadron momentum, i.e., in the helicity basis.
The transverse directions can be chosen as the normal of the lepton-hadron plane
(defined by the momenta of the hadron $h$ and the electron $e^-$), i.e., the $y$-direction,
or that of the hadron-jet plane (defined by the momenta of the hadron $h$ and the $\bar q$).
The expressions of these different components of polarizations in terms of the structure functions can easily be derived
and take exactly the same form as those for $e^+e^-\to V\pi X$ given in \cite{Chen:2016moq}.
We do not repeat them here.

If we integrate over $d^2k'_\perp$, we obtain the result for the inclusive process $e^+e^-\to hX$, i.e.,
\begin{align}
 & \frac{E_p d\sigma^{(in)}}{d^3 p}=\frac{\alpha^2 \chi}{ s^2} \Bigl[
 \mathcal{F}^{(in)}_U +\lambda_h \tilde{\mathcal{F}}^{(in)}_L
+|S_T|(\mathcal{F}^{(in)}_T+ \tilde{\mathcal{F}}^{(in)}_T) \nonumber\\
&~~~+S_{LL}\mathcal{F}^{(in)}_{LL}+|S_{LT}|(\mathcal{F}^{(in)}_{LT}+ \tilde{\mathcal{F}}^{(in)}_{LT})
+|S_{TT}|(\mathcal{F}^{(in)}_{TT}+ \tilde{\mathcal{F}}^{(in)}_{TT})\Bigr], \label{eq:csinclusive} \\
&{\cal F}_{U}^{(in)} =(1+\cos^2\theta) F_{U1}^{(in)}+  \sin^2\theta F_{U2}^{(in)} + \cos\theta F_{U3}^{(in)},\label{eq:FUin} \\
&\tilde{\cal F}_{L}^{(in)} = (1+\cos^2\theta) \tilde F_{L1}^{(in)}+  \sin^2\theta \tilde F_{L2}^{(in)} + \cos\theta \tilde F_{L3}^{(in)}, \label{eq:tFLin} \\
&{\cal F}_{LL}^{(in)}= (1+\cos^2\theta) F_{LL 1}^{(in)}+  \sin^2\theta F_{LL2}^{(in)} + \cos\theta F_{LL3}^{(in)}, \label{eq:FLLin} \\
&{\cal F}_{T}^{(in)}= \sin\varphi_S ( \sin\theta F_{T1}^{(in)\sin\varphi_S} + \sin2\theta F_{T2}^{(in)\sin\varphi_S} ), \label{eq:FTin}\\
&\tilde{\cal F}_{T}^{(in)}= \cos\varphi_S ( \sin\theta \tilde F_{T1}^{(in)\cos\varphi_S} + \sin2\theta \tilde F_{T2}^{(in)\cos\varphi_S} ), \label{eq:tFTin} \\
&{\cal F}_{LT}^{(in)}= \cos\varphi_{LT} ( \sin\theta F_{LT1}^{(in)\cos\varphi_{LT}} + \sin2\theta F_{LT2}^{(in)\cos\varphi_{LT}} ), \label{eq:FLTin}\\
&\tilde{\cal F}_{LT}^{(in)} = \sin\varphi_{LT} ( \sin\theta \tilde F_{LT1}^{(in)\sin\varphi_{LT}} + \sin2\theta \tilde F_{LT2}^{(in)\sin\varphi_{LT}} ), \label{eq:tFLTin} \\
&{\cal F}_{TT}^{(in)}= \cos2\varphi_{TT} \sin^2\theta F_{TT}^{(in)\cos2\varphi_{TT}}, \label{eq:FTTin} \\
&\tilde{\cal F}_{TT}^{(in)}= \sin2\varphi_{TT} \sin^2\theta \tilde F_{TT}^{(in)\sin2\varphi_{TT}}. \label{eq:tFTTin}
\end{align}
We have in total 19 inclusive structure functions
and they are just equal to the semi-inclusive counterparts integrated over $d^2k'_\perp/(2\pi)^2$.

\subsection{Hadronic tensor in the QCD parton model}

\begin{figure}
 \includegraphics[width=8cm]{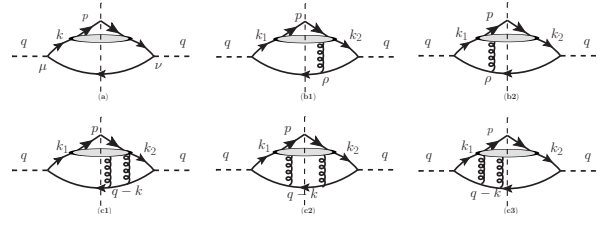}
  \caption{The first few diagrams as examples of the considered diagram series with exchange of $j$-gluon(s) and different cuts.
 We see (a) $j=0$, (b1) $j=1$ and left cut, (b2) $j=1$ and right cut,
  (c1)  $j=2$ and left cut, (c2) $j=2$ and middle cut,  and (c3) $j=2$ and right cut, respectively.} \label{fig:Feyndiagram}
\end{figure}

In the QCD parton model, at the tree level of pQCD, we need to consider the series of diagrams illustrated in Fig.~\ref{fig:Feyndiagram}
where diagrams with exchange of $j$ gluon(s) ($j=0,1,2, \cdots$) are included.
After the collinear expansion, the TMD semi-inclusive hadronic tensor is obtained as
\begin{align}
&W_{\mu\nu}(q,p,S,k'_\perp)=\sum_{j,c}\tilde W^{(j,c)}_{\mu\nu}(q,p,S,k'_\perp),\label{eq:Wsum}
\end{align}
where $c$ denotes different cuts.
The  $\tilde W^{(j,c)}_{\mu\nu}$ is a trace of the collinear-expanded hard part and gauge invariant quark-$j$-gluon-quark correlator and can be simplified to~\cite{Wei:2014pma}
\begin{align}
   & \tilde W^{(0)}_{\mu\nu}
   = \frac{1}{2} \mathrm{Tr}\big[\hat h^{(0)}_{\mu\nu} \hat\Xi^{(0)}\big],\label{eq:tW0}\\
   & \tilde W^{(1,L)}_{\mu\nu}
   =-\frac{1}{4(p\cdot q)} \mathrm{Tr}\big[\hat h^{(1)\rho}_{\mu\nu}\hat\Xi^{(1)}_{\rho} \big],\label{eq:tW1L}\\
   & \tilde W^{(2,L)}_{\mu\nu}
   =\frac{1}{4(p\cdot q)^2} \mathrm{Tr}\big[ \hat N_{\mu\nu}^{(2)\rho\sigma} \hat\Xi^{(2)}_{\rho\sigma} +\hat h^{(1)\rho}_{\mu\nu}\hat\Xi^{(2\prime)}_{\rho} \big],\label{eq:tW2L} \\
   & \tilde W^{(2,M)}_{\mu\nu}
   =\frac{1}{4(p\cdot q)^2} \mathrm{Tr}\big[\hat h^{(2)\rho\sigma}_{\mu\nu} \hat\Xi^{(2,M)}_{\rho\sigma} \big],\label{eq:tW2M}
\end{align}
where the hard parts are given by
\begin{align}
  &\hat h^{(0)}_{\mu\nu} =\frac{1}{p^+}\Gamma^q_\mu \slashed n \Gamma^q_\nu,\label{eq:h0}\\
  &\hat h^{(1)\rho}_{\mu\nu} =\Gamma^q_\mu \slashed n \gamma_\perp^\rho \slashed{\bar n}\Gamma^q_\nu,\label{eq:h1}\\
  &\hat h^{(2)\rho\sigma}_{\mu\nu} =\frac{p^+}{2}\Gamma^q_\mu\slashed{\bar n}\gamma_\perp^\rho\slashed n\gamma_\perp^\sigma\slashed{\bar n}\Gamma^q_\nu, \label{eq:h2} \\
  &\hat N_{\mu\nu}^{(2)\rho\sigma} =q^-\Gamma^q_\mu\gamma_\perp^\rho\slashed n \gamma_\perp^\sigma \Gamma^q_\nu.\label{eq:N2}
\end{align}
All the quark-quark and quark-$j$-gluon-quark correlators involved are functions of one parton momentum and the hadron momentum and spin,
i.e., ($z$, $k_\perp$, $p$, $S$), and are given by
\begin{align}
  \hat\Xi^{(0)}  = & \sum_X \int \frac{ p^+ d\xi^- d^2\xi_\perp}{2\pi} e^{-ip^+\xi^-/z+ik_\perp \cdot  \xi_\perp} \langle 0|\mathcal{L}^\dag(0,\infty) \nonumber\\
  &\times \psi(0)|hX \rangle \langle hX|\bar \psi(\xi)\mathcal{L}(\xi,\infty) |0 \rangle,\label{eq:Xi0}\\
  \hat\Xi^{(1)}_\rho  = & \sum_X \int \frac{ p^+ d\xi^- d^2\xi_\perp}{2\pi} e^{-ip^+\xi^-/z+i k_\perp \cdot \xi_\perp}    \langle 0|\mathcal{L}^\dag(0,\infty) \nonumber\\
  &\times D_{\perp\rho}(0)\psi(0)|hX \rangle \langle hX|\bar \psi(\xi)\mathcal{L}(\xi,\infty) |0 \rangle,\label{eq:Xi1}\\
 \hat\Xi^{(2)}_{\rho\sigma} = & \sum_X \int \frac{ p^+ d\xi^- d^2\xi_\perp}{2\pi}\int_0^\infty ip^+d\eta^- e^{-ip^+\xi^-/z+ik_\perp \cdot  \xi_\perp} \nonumber\\
  &\times  \langle 0|\mathcal{L}^\dag(\eta,\infty)D_{\perp\rho}(\eta)D_{\perp\sigma}(\eta) \mathcal{L}^\dag(0,\eta) \nonumber\\
  &\times\psi(0)|hX \rangle \langle hX|\bar \psi(\xi)\mathcal{L}(\xi,\infty) |0 \rangle,\label{eq:Xi2} \\
  \hat\Xi^{(2\prime)}_\rho = & \sum_X \int \frac{ p^+ d\xi^- d^2\xi_\perp}{2\pi} e^{-ip^+\xi^-/z+i k_\perp \cdot \xi_\perp}p^\sigma \langle 0|\mathcal{L}^\dag(0,\infty) \nonumber\\
  &\times D_{\perp\rho}(0)D_\sigma(0)\psi(0)|hX \rangle \langle hX|\bar \psi(\xi)\mathcal{L}(\xi,\infty) |0 \rangle,\label{eq:Xi2prime}\\
 \hat\Xi^{(2,M)}_{\rho\sigma} = & \sum_X \int \frac{ p^+ d\xi^- d^2\xi_\perp}{2\pi} e^{-ip^+\xi^-/z+i k_\perp \cdot \xi_\perp} \langle 0|\mathcal{L}^\dag(0,\infty) \nonumber\\
  &\times D_{\perp\rho}\psi(0)|hX \rangle \langle hX|\bar \psi(\xi)D_{\perp\sigma}(\xi)\mathcal{L}(\xi,\infty) |0 \rangle,\label{eq:Xi2M}
\end{align}
where $D_\rho=-i\partial_\rho+gA_\rho$, and $\mathcal{L}(0,y)$ is the gauge link.
As a convention, the argument $\xi$ in the quark filed operator $\psi$ and gauge link represents $(0,\xi^-,\vec\xi_\perp)$.
We note that the leading power contribution of $\tilde W^{(j)}_{\mu\nu}$ is twist-$(j+2)$.
However, because of the factor $p^\sigma$ in the definition of $\hat\Xi^{(2\prime)}_\rho$ given by Eq.~(\ref{eq:Xi2prime}),
the second term in Eq.~(\ref{eq:tW2L}) has no contribution up to twist-4.
The leading power contribution of this term is twist-5.

\subsection{Decompositions of the quark-$j$-gluon-quark correlator}\label{sec:decomp}

In $e^+e^-\to h\bar q X$, only the chiral even FFs are involved.
We only need to consider the $\gamma^\alpha$- and the $\gamma^5\gamma^\alpha$-term in the decomposition
of the correlators in terms of the $\Gamma$-matrices such as $\hat\Xi^{(0)} =\Xi_\alpha^{(0)}\gamma^\alpha+\tilde\Xi_\alpha^{(0)} \gamma^5\gamma^\alpha+\cdots$.
We write down all the twist-4 terms in the decomposition of these correlators in the following.
For $\hat\Xi^{(0)}$, they are given by
\begin{align}
  z\Xi^{(0)}_\alpha =&\frac{M^2}{p^+}n_\alpha \Big(D_3 - \frac{\varepsilon_\perp^{kS}}{M}D^\perp_{3T}+ S_{LL}D_{3LL}\nonumber\\
  &+\frac{k_\perp \cdot S_{LT}}{M}D_{3LT}^{\perp}+\frac{S_{TT}^{kk}}{M^2}D_{3TT}^{\perp} \Big), \label{eq:Xi0decomp}\\
  z\tilde\Xi^{(0)}_\alpha =&-\frac{M^2}{p^+}n_\alpha \Big(\lambda_hG_{3L} - \frac{k_\perp\cdot S_T}{M}G^\perp_{3T}\nonumber\\
  &+\frac{\varepsilon_\perp^{kS_{LT}}}{M}G_{3LT}^{\perp}+\frac{S_{TT}^{\tilde k k}}{M^2}G_{3TT}^{\perp} \Big). \label{eq:tXi0decomp}
\end{align}
Here, as in~\cite{Chen:2016moq}, $D$'s and $G$'s represent the $\gamma^\alpha$- and $\gamma^5\gamma^\alpha$-type FFs, respectively.
The digit $j$ in the subscript denotes twist-$(j+1)$;
the capital letter such as $T, L$, or $LL$ denotes the hadron polarization.
There are in total nine twist-4 chiral even FFs defined via $\hat\Xi^{(0)}$.

For $\hat\Xi^{(1)}_\rho $, the chiral even parts are
\begin{align}
  z\Xi^{(1)}_{\rho\alpha} & = M^2 g_{\perp\rho\alpha} \Big(D_{3d}- \frac{\varepsilon_\perp^{kS}}{M} D_{3dT}^\perp +S_{LL}D_{3dLL} \nonumber\\
  &\hspace{10mm}+\frac{k_\perp \cdot S_{LT}}{M}D_{3dLT}^{\perp}+\frac{S_{TT}^{kk}}{M^2}D_{3dTT}^{\perp}\Big) \nonumber\\
  &+ k_{\perp\langle\rho}k_{\perp\alpha\rangle} \Big(D_{3d}^\perp+\frac{\varepsilon_\perp^{kS}}{M}D_{3dT}^{\perp 2}+ S_{LL}D_{3dLL}^\perp \nonumber\\
  &\hspace{10mm}+\frac{k_\perp \cdot S_{LT}}{M}D_{3dLT}^{\perp2} +\frac{S_{TT}^{kk}}{M^2}D_{3dTT}^{\perp2}\Big) \nonumber\\
  & + i M^2\varepsilon_{\perp\rho\alpha} \Big(\lambda_h D_{3dL}-\frac{k_\perp\cdot S_T}{M} D_{3dT}^{\perp 3} \nonumber\\
  &\hspace{10mm} + \frac{\varepsilon_\perp^{kS_{LT}}}{M}D_{3dLT}^{\perp3}+\frac{S_{TT}^{\tilde k k}}{M^2}D_{3dTT}^{\perp3}  \Big)\nonumber\\
  & + \frac{1}{2}k_{\perp\{\rho} \tilde k_{\perp \alpha \}} \Big(\lambda_h D_{3dL}^\perp+\frac{k_\perp\cdot S_T}{M}D_{3dT}^{\perp 4} \nonumber\\
  &\hspace{10mm} +\frac{\varepsilon_\perp^{kS_{LT}}}{M}D_{3dLT}^{\perp4}+\frac{S_{TT}^{\tilde k k}}{M^2}D_{3dTT}^{\perp4}  \Big), \label{eq:Xi1decomp} \\
  z\tilde\Xi^{(1)}_{\rho\alpha} & = iM^2\varepsilon_{\perp\rho\alpha} \Big(G_{3d} - \frac{\varepsilon_\perp^{kS}}{M}G_{3dT}^\perp +S_{LL}G_{3dLL} \nonumber\\
  &\hspace{10mm} +\frac{k_\perp \cdot S_{LT}}{M}G_{3dLT}^{\perp}+\frac{S_{TT}^{kk}}{M^2}G_{3dTT}^{\perp}\Big) \nonumber\\
  & +  \frac{i}{2}k_{\perp\{\rho} \tilde k_{\perp \alpha \}} \Big(G_{3d}^\perp + \frac{\varepsilon_\perp^{kS}}{M}  G_{3dT}^{\perp 2} +S_{LL}G_{3dLL}^\perp\nonumber\\
  &\hspace{10mm}+\frac{k_\perp \cdot S_{LT}}{M}G_{3dLT}^{\perp2}+\frac{S_{TT}^{kk}}{M^2}G_{3dTT}^{\perp2}\Big) \nonumber\\
  &+M^2 g_{\perp\rho\alpha}\Big(\lambda_h G_{3dL} - \frac{k_\perp\cdot S_T}{M} G_{3dT}^{\perp 3} \nonumber\\
  &\hspace{10mm} +\frac{\varepsilon_\perp^{kS_{LT}}}{M}G_{3dLT}^{\perp3}+\frac{S_{TT}^{\tilde k k}}{M^2}G_{3dTT}^{\perp3}  \Big)\nonumber\\
  &+ i k_{\perp\langle\rho}k_{\perp\alpha\rangle} \Big(\lambda_h G_{3dL}^\perp+\frac{k_\perp\cdot S_T}{M} G_{3dT}^{\perp 4} \nonumber\\
  &\hspace{10mm} + \frac{\varepsilon_\perp^{kS_{LT}}}{M}G_{3dLT}^{\perp4}+\frac{S_{TT}^{\tilde k k}}{M^2}G_{3dTT}^{\perp4}  \Big),\label{eq:tXidecomp}
\end{align}
where $k_{\perp\langle\rho}k_{\perp\alpha\rangle}\equiv k_{\perp\rho}k_{\perp\alpha}-{k_\perp^2}g_{\perp\rho\alpha}/2$ and $g_{\perp\rho\alpha}$ 
is defined as  $ g_{\perp\rho\alpha}=g_{\rho\alpha}-\bar n_\rho n_\alpha-\bar n_\alpha n_\rho$.
Here we add a subscript $d$ to denote FFs defined via $\hat\Xi^{(1)}_\rho$.

Up to twist-4, we only need the leading power contributions from  $\hat\Xi^{(2)}_{\rho\sigma}$ and $\hat\Xi^{(2,M)}_{\rho\sigma}$.
For  the chiral even part, we need only the $\bar n_\alpha$-terms.
They are given by
\begin{align}
 z\Xi^{(2)}_{\rho\sigma\alpha} &= p^+\bar n_\alpha \Big[
   M^2 g_{\perp\rho\sigma} \Big(D_{3dd} - \frac{\varepsilon_\perp^{kS} }{M} D_{3ddT}^{\perp} \nonumber\\
   &\hspace{5mm} +S_{LL}D_{3ddLL}+\frac{k_\perp \cdot S_{LT}}{M}D_{3ddLT}^{\perp}+\frac{S_{TT}^{kk}}{M^2}D_{3ddTT}^{\perp}\Big)     \nonumber\\
   &+ k_{\perp\langle\rho}k_{\perp\sigma\rangle}
   \Big(D_{3dd}^{\perp} + \frac{\varepsilon_\perp^{kS}}{M}  D_{3ddT}^{\perp 2}  \nonumber\\
   &\hspace{5mm} +S_{LL}D_{3ddLL}^\perp - \frac{k_\perp \cdot S_{LT}}{M}D_{3ddLT}^{\perp2} -\frac{S_{TT}^{kk}}{M^2}D_{3ddTT}^{\perp2}\Big)  \nonumber\\
   &+ iM^2\varepsilon_{\perp\rho\sigma} \Big(\lambda_h D_{3ddL}-\frac{k_\perp\cdot S_T}{M} D_{3ddT}^{\perp 3}  \nonumber\\
   &\hspace{5mm} -\frac{\varepsilon_\perp^{kS_{LT}}}{M}D_{3ddLT}^{\perp3}-\frac{S_{TT}^{\tilde k k}}{M^2}D_{3ddTT}^{\perp3} \Big)  \nonumber\\
   &+ \frac{1}{2}k_{\perp\{\rho} \tilde k_{\perp \sigma \}}  \Big(\lambda_h D_{3ddL}^{\perp}+\frac{k_\perp \cdot S_T}{M} D_{3ddT}^{\perp 4}   \nonumber\\
   &\hspace{5mm} +\frac{\varepsilon_\perp^{kS_{LT}}}{M}D_{3ddLT}^{\perp4}
    + \frac{S_{TT}^{\tilde k k}}{M^2}D_{3ddTT}^{\perp4} \Big)  \Big], \label{eq:Xi2decomp} \\
 z\tilde\Xi^{(2)}_{\rho\sigma\alpha} & = p^+\bar n_\alpha \Big[
    iM^2 \varepsilon_{\perp\rho\sigma} \Big(G_{3dd} - \frac{\varepsilon_\perp^{kS}}{M} G_{3ddT}^{\perp} \nonumber\\
  &\hspace{5mm} +S_{LL}G_{3ddLL}+\frac{k_\perp \cdot S_{LT}}{M}G_{3ddLT}^{\perp}+\frac{S_{TT}^{kk}}{M^2}G_{3ddTT}^{\perp}\Big)  \nonumber\\
  &+  \frac{1}{2}k_{\perp\{\rho} \tilde k_{\perp \sigma \}} \Big(G_{3dd}^{\perp} + \frac{\varepsilon_\perp^{kS}}{M} G_{3ddT}^{\perp 2} \nonumber\\
  &\hspace{5mm} +S_{LL}G_{3ddLL}^\perp - \frac{k_\perp \cdot S_{LT}}{M}G_{3ddLT}^{\perp2}-\frac{S_{TT}^{kk}}{M^2}G_{3ddTT}^{\perp2}\Big) \nonumber\\
  &+  M^2 g_{\perp\rho\sigma} \Big(\lambda_h G_{3ddL} - \frac{k_\perp\cdot S_T}{M} G_{3ddT}^{\perp 3} \nonumber\\
  &\hspace{5mm} -\frac{\varepsilon_\perp^{kS_{LT}}}{M}G_{3ddLT}^{\perp3}-\frac{S_{TT}^{\tilde k k}}{M^2}G_{3ddTT}^{\perp3} \Big) \nonumber\\
  &+ k_{\perp\langle\rho}k_{\perp\sigma\rangle} \Big(\lambda_h G_{3ddL}^{\perp} + \frac{k_\perp\cdot S_T}{M} G_{3ddT}^{\perp 4}   \nonumber\\
  &\hspace{5mm} +\frac{\varepsilon_\perp^{kS_{LT}}}{M}G_{3ddLT}^{\perp4}+\frac{S_{TT}^{\tilde k k}}{M^2}G_{3ddTT}^{\perp4} \Big) \Big], \label{eq:tXi2decomp}
\end{align}
where  we use $dd$ in the subscript to denote FFs defined via $\hat\Xi^{(2)}_{\rho\sigma}$.
The decomposition of $\hat\Xi^{(2,M)}_{\rho\sigma} $ takes exactly the same form as that of $\hat\Xi^{(2)}_{\rho\sigma}$.
We just add an additional superscript $M$ to distinguish them from each other and omit the equations here.

From Eqs.~(\ref{eq:Xi0decomp})-(\ref{eq:tXi2decomp}), we see that for the twist-4 parts,
the decomposition of $\Xi$ and that of $\tilde \Xi$ have exact one to one correspondence.
For each $D_3$, there is a $G_3$ corresponding to it.
They always appear in pairs.
Because of the Hermiticity of $\hat\Xi^{(0)}$ and $\hat\Xi^{(2,M)}_{\rho\sigma}$, the FFs defined via them are real.
For those defined via $\hat\Xi^{(1)}_{\rho}$ and $\hat\Xi^{(2)}_{\rho\sigma}$, there is no such constraint so that they can be complex.

\subsection{Relationships derived from the QCD equation of motion}

From the QCD equation of motion, $\gamma \cdot D \psi=0$,
we relate the quark-$j$-gluon-quark correlators to the quark-quark correlator.
For the two transverse components $\Xi_{\perp}^{(0)\rho}$ and $\tilde\Xi_{\perp}^{(0)\rho}$, we have
\begin{align}
  k^+\Xi_{\perp}^{(0)\rho}&=-g_\perp^{\rho\sigma} \mathrm{Re}\Xi^{(1)}_{\sigma +}-\varepsilon^{\rho\sigma}_{\perp }\mathrm{ Im} \tilde \Xi^{(1)}_{\sigma +},\label{eq:perpeom}\\
  k^+\tilde \Xi_{\perp}^{(0)\rho}&=-g_\perp^{\rho\sigma} \mathrm{Re}\tilde \Xi^{(1)}_{\sigma +}-\varepsilon^{\rho\sigma}_{\perp }\mathrm{ Im} \Xi^{(1)}_{\sigma +}. \label{eq:tperpeom}
\end{align}
Eqs.~(\ref{eq:perpeom}) and (\ref{eq:tperpeom}) lead to a set of relationships between twist-3 FFs given in the unified form~\cite{Chen:2016moq}
\begin{align}
  D_{S}^K - iG_{S}^K =-z(D_{dS}^K-G_{dS}^K),\label{eq:t3eom}
\end{align}
where $S=$null, $L, T, LL, LT$ or $TT$ and $K=$null, $\perp$ or $\prime\hspace{-0.17cm}\perp$ whenever applicable~\cite{footnote}.
Similarly, for the minus components of $\Xi^{(0)}_\alpha$ and $\tilde\Xi^{(0)}_\alpha$, we have
\begin{align}
  2k^{+2}\Xi^{(0)}_- &=k^+\Big(g_\perp^{\rho\sigma} \Xi^{(1)}_{\rho\sigma } + i\varepsilon_\perp^{\rho\sigma}\tilde \Xi^{(1)}_{\rho\sigma }\Big) \nonumber\\
  & =- g_\perp^{\rho\sigma} \Xi^{(2,M)}_{\rho\sigma +}+i\varepsilon_\perp^{\rho\sigma}\tilde \Xi^{(2,M)}_{\rho\sigma +},\label{eq:minuseom}\\
  2k^{+2}\tilde\Xi^{(0)}_- &=k^+\Big(g_\perp^{\rho\sigma}\tilde \Xi^{(1)}_{\rho\sigma } + i\varepsilon_\perp^{\rho\sigma} \Xi^{(1)}_{\rho\sigma }\Big) \nonumber\\
  & =- g_\perp^{\rho\sigma}\tilde \Xi^{(2,M)}_{\rho\sigma +}+i\varepsilon_\perp^{\rho\sigma}\Xi^{(2,M)}_{\rho\sigma +}. \label{eq:tminuseom}
\end{align}
From Eqs.~(\ref{eq:minuseom}) and (\ref{eq:tminuseom}), we obtain a set of relationships between twist-4 FFs
defined via $\hat\Xi^{(0)}$, $\hat\Xi^{(1)}$ and $\hat\Xi^{(2,M)}$.
For longitudinal components, we have
\begin{align}
&D_{3}  = z D_{-3d} = - z^2D_{+3dd}^{M}, \label{eq:t4eomD} \\
&D_{3LL}  = z D_{-3dLL} = - z^2D_{+3ddLL}^{M}, \label{eq:t4eomDLL} \\
&G_{3L}  = z D_{-3dL} = z^2D_{+3ddL}^{M}, \label{eq:t4eomGL}
\end{align}
where $D_{\pm}\equiv D\pm G$ such as $D_{-3d}\equiv D_{3d}-G_{3d}$ and so on.
For the transverse components, we have
\begin{align}
&D_{3S}^\perp  = z D_{-3dS}^\perp = - z^2D_{+3ddS}^{M\perp}, \label{eq:t4eomDt}\\
&G_{3S}^\perp= z D_{-3dS}^{\perp3} = -\eta_Sz^2D_{+3ddS}^{M\perp3}, \label{eq:t4eomGt}
\end{align}
where $S=T, LT$ or $TT$ represents the transverse components; $\eta_S$ represents a sign that takes $-1$ for $S=T$ and $+1$ for $S=LT$ or $TT$ as well as in Eqs.~(\ref{eq:g0D3dTperp}) and (\ref{eq:g0G3dTLperp}) .

We note in particular that Eqs.~(\ref{eq:t4eomD})-(\ref{eq:t4eomGt}) represent totally 27 equations and
can be used to eliminate those twist-4 TMD FFs that are not independent in parton model results for cross section.

\subsection{Relationships between twist-4 and leading twist FFs at g=0 }

The higher twist FFs defined in Sec.~\ref{sec:decomp} are new and much involved. Currently there is not much data available.
If we neglect the multiple gluon scattering, i.e., set $g=0$, we obtain a set of equations relating them to the leading twist counterparts.
These relationships could be helpful in understanding the properties of these higher twist FFs in particular at the present stage
when few data are available. We give these relationships in this subsection.

By putting $g=0$ into Eqs.~(\ref{eq:Xi0})-(\ref{eq:Xi2M}), we obtain relationships such as
$\hat\Xi^{(1)}_\rho|_{g=0}=-k_{\perp\rho} \hat\Xi^{(0)}|_{g=0}$,
$\hat\Xi^{(2,M)}_{\rho\sigma}|_{g=0}=k_{\perp\rho} k_{\perp\sigma} \hat\Xi^{(0)}|_{g=0}$,  $(\hat\Xi^{(2)}_{\rho\sigma}+\gamma^0\hat\Xi^{(2)\dag}_{\sigma\rho}\gamma^0)|_{g=0} =z^2k_{\perp\rho}k_{\sigma\perp}(\partial\hat \Xi^{(0)}/\partial z)|_{g=0}$.
Together with the QCD equation of motion, we obtain the relationships between the leading twist FFs and twist-4 FFs in the following.
For the twist-4 FFs defined via $\hat\Xi^{(1)}_\rho$, we obtain that for the longitudinal components
\begin{align}
  & D_{3d} = \frac{k_\perp^2}{2M^2} D_{3d}^\perp = \frac{1}{z} D_3 = -\frac{k_\perp^2}{2M^2} zD_1, \label{eq:g0D3d}\\
  & D_{3dLL} = \frac{k_\perp^2}{2M^2} D_{3dLL}^\perp = \frac{1}{z} D_{3LL} = -\frac{k_\perp^2}{2M^2} z D_{1LL}, \label{eq:g0D3dLL}\\
  & G_{3dL} = i \frac{k_\perp^2}{2M^2} G_{3dL}^\perp = \frac{1}{z} G_{3L} = -\frac{k_\perp^2}{2M^2} z G_{1L}, \label{eq:g0G3dL}
\end{align}
and for the transverse components
\begin{align}
  & D_{3dS}^\perp = \eta_S \frac{k_\perp^2}{2M^2} D_{3dS}^{\perp 2} = \frac{1}{z} D_{3S}^\perp = -\frac{k_\perp^2}{2M^2} z D_{1S}^\perp, \label{eq:g0D3dTperp}\\
  & G_{3dS}^{\perp 3} = i \eta_S \frac{k_\perp^2}{2M^2} G_{3dS}^{\perp 4} = \frac{1}{z} G_{3S}^\perp = \eta_S \frac{k_\perp^2}{2M^2} zG_{1S}^\perp, \label{eq:g0G3dTLperp}
\end{align}
where $S=T, LT$ or $TT$.

For those defined via $\hat\Xi^{(2)}_{\rho\sigma}$, we have, for the longitudinal components,
\begin{align}
  &  \mathrm{Re}D_{3dd} = \frac{k_\perp^2}{2M^2}  \mathrm{Re}D_{3dd}^{\perp } = z^2\frac{k_\perp^2}{4M^2}\frac{\partial}{\partial z}D_1, \label{eq:g0D3dd}\\
  &  \mathrm{Re}G_{3ddL} = \frac{k_\perp^2}{2M^2}  \mathrm{Re}G_{3ddL}^{\perp } = -z^2\frac{k_\perp^2}{4M^2}\frac{\partial}{\partial z}G_{1L}, \label{eq:g0G3ddL}\\
  &  \mathrm{Re}D_{3ddLL} = \frac{k_\perp^2}{2M^2}  \mathrm{Re}D_{3ddLL}^{\perp } = z^2\frac{k_\perp^2}{4M^2}\frac{\partial}{\partial z}D_{1LL}, \label{eq:g0D3ddLL}
  \end{align}
  and for the transverse components
  \begin{align}
  &  \mathrm{Re}D_{3ddS}^\perp = - \frac{k_\perp^2}{2M^2} \mathrm{Re} D_{3ddS}^{\perp 2} = z^2\frac{k_\perp^2}{4M^2}\frac{\partial}{\partial z}D_{1S}^\perp, \label{eq:g0D3ddST}\\
  &  \mathrm{Re}G_{3ddS}^{\perp 3} = - \frac{k_\perp^2}{2M^2} \mathrm{Re}G_{3ddS}^{\perp 4} = - z^2\frac{k_\perp^2}{4M^2}\frac{\partial}{\partial z}G_{1S}^\perp, \label{eq:g0G3ddST}
  \end{align}
where $S=T, LT$ or $TT$.
All others twist-4 FFs vanish and also time-reversal invariance demands $D_{1T}^\perp=0$ in this case.

\section{The complete twist-4 results}\label{sec:results}

By substituting Eqs.~(\ref{eq:h0})-(\ref{eq:N2}) and (\ref{eq:Xi0decomp})-(\ref{eq:tXi2decomp}) into
Eqs.~(\ref{eq:tW0})-(\ref{eq:tW2M}), carrying out the traces, we obtain the hadronic tensor results at twist-4.
Making the Lorentz contraction of the hadronic tensor with the leptonic tensor, we obtain the cross section up to twist-4.
We compare the results with the general form of the cross section given by Eqs.~(\ref{eq:cs})-(\ref{eq:tWTT}) and obtain
the results for the structure functions in terms of gauge invariant FFs.
We present the complete results up to twist-4 in this section.
For comparison, we also show the corresponding results at leading twist and twist-3.
They can be found, e.g., in~\cite{Wei:2014pma}.
There are also contributions from the four-quark correlators at twist-4.
We present them in Sec.~\ref{sec:4q}.
We also show the results for the inclusive processes. \\

\subsection{The hadronic tensor at twist-4}
The hadronic tensor up to twist-3 obtained using the formalism in Sec.~\ref{sec:formalism} has been presented in e.g.~\cite{Wei:2014pma}.
We show only the twist-4 part obtained by substituting Eqs.~(\ref{eq:h0})-(\ref{eq:N2}) and Eqs.~(\ref{eq:Xi0decomp})-(\ref{eq:tXi2decomp}) into Eqs.~(\ref{eq:tW0})-(\ref{eq:tW2M}).

For the contributions from $\tilde W^{(0)}_{\mu\nu}$, we use
\begin{align}
 & \mathrm{Tr}\big[\hat h^{(0)}_{\mu\nu} \gamma^\alpha\big]
 =-\frac{4}{p^+}\Big[c^q_1(g_{\mu\nu} n^\alpha-g^\alpha_{\{\mu} n_{\nu\}})+ic^q_3\varepsilon_{\mu\nu}^{~~\alpha n} \Big], \label{eq:h0e}\\
 & \mathrm{Tr}\big[\hat h^{(0)}_{\mu\nu}\gamma^5\gamma^\alpha\big]
 =\frac{4}{p^+}\Big[c^q_3(g_{\mu\nu} n^\alpha-g^\alpha_{\{\mu} n_{\nu\}})+ic^q_1\varepsilon_{\mu\nu}^{~~\alpha n} \Big], \label{eq:h0o}
\end{align}
and obtain the twist-4 part of $\tilde W^{(0)}_{\mu\nu}$ as given by
\begin{align}
 \tilde W^{(0)}_{t4\mu\nu} =& \frac{4M^2n_\mu n_\nu }{z(p^+)^2}  \Big[ c^q_1  \Big(D_3 - \frac{\varepsilon_\perp^{kS}}{M}D^\perp_{3T} + S_{LL}D_{3LL} \nonumber\\
  & + \frac{k_\perp \cdot S_{LT}}{M}D_{3LT}^{\perp}+\frac{S_{TT}^{kk}}{M^2}D_{3TT}^{\perp} \Big)  +c^q_3 \Big(\lambda_h G_{3L}  \nonumber\\
  & - \frac{k_\perp\cdot S_T}{M}G^\perp_{3T} +\frac{\varepsilon_\perp^{kS_{LT}}}{M}G_{3LT}^{\perp}+\frac{S_{TT}^{\tilde k k}}{M^2}G_{3TT}^{\perp} \Big)\Big],\label{eq:w0}
\end{align}
where we use a subscript $t4$ to denote the twist-4 part only.

For $\tilde W^{(1)}_{\mu\nu}$, we have contributions from $\tilde W^{(1,L)}_{\mu\nu}$ and $\tilde W^{(1,R)}_{\mu\nu}=\tilde W^{(1,L)*}_{\nu\mu}$. We calculate
\begin{align}
 \mathrm{Tr}\big[\hat h^{(1)\rho}_{\mu\nu} \gamma^\alpha\big]
 &= 4c^q_1\Big[2n_\mu \bar n_\nu g_{\perp}^{\rho\alpha} +g_{\perp\mu\nu}g_{\perp}^{\rho\alpha}-g_{\perp\mu}^{~~~\{\rho}g_{\perp\nu}^{~~~\alpha\}}\Big]\nonumber\\
  &- 4ic^q_3\Big[2n_\mu \bar n_\nu \varepsilon_{\perp}^{\rho\alpha}+g_{\perp\mu}^{~~~\rho}\varepsilon_{\perp\nu}^{~~~\alpha}+g_{\perp\nu}^{~~\alpha}\varepsilon_{\perp\mu} ^{~~~\rho}\Big],\label{eq:h1oe}\\
 \mathrm{Tr}\big[\hat h^{(1)\rho}_{\mu\nu}\gamma^5 \gamma^\alpha\big]
  &=4ic^q_1\Big[2n_\mu \bar n_\nu \varepsilon_{\perp}^{\rho\alpha}+g_{\perp\mu}^{~~~\rho}\varepsilon_{\perp\nu}^{~~~\alpha}+g_{\perp\nu}^{~~~\alpha}\varepsilon_{\perp\mu} ^{~~~\rho}\Big]\nonumber\\
  &-4c^q_3\Big[2n_\mu \bar n_\nu g_{\perp}^{\rho\alpha} +g_{\perp\mu\nu}g_{\perp}^{\rho\alpha}-g_{\perp\mu}^{~~~\{\rho}g_{\perp\nu}^{~~~\alpha\}}\Big], \label{eq:h1oo}
\end{align}
and obtain
\begin{widetext}
\begin{align}
  \tilde W^{(1,L)}_{t4\mu\nu} =& -\frac{4M^2}{z(p\cdot q)}n_\mu \bar n_\nu  \Big[ c^q_1  \Big(D_{-3d}- \frac{\varepsilon_\perp^{kS}}{M} D_{-3dT}^\perp  +S_{LL}D_{-3dLL} 
  +\frac{k_\perp \cdot S_{LT}}{M}D_{-3dLT}^{\perp}+\frac{S_{TT}^{kk}}{M^2}D_{-3dTT}^{\perp}\Big)  \nonumber\\
  &\phantom{XXXXXX} + c^q_3 \Big(\lambda_h D_{-3dL} -\frac{k_\perp\cdot S_T}{M} D_{-3dT}^{\perp 3} 
  +\frac{\varepsilon_\perp^{kS_{LT}}}{M}D_{-3dLT}^{\perp3}+\frac{S_{TT}^{\tilde k k}}{M^2}D_{-3dTT}^{\perp3}  \Big) \Big] \nonumber\\
  +& \frac{2}{z(p\cdot q)}k_{\perp\langle\mu}k_{\perp\nu\rangle} \Big[c^q_1\Big(D_{-3d}^\perp+ \frac{\varepsilon_\perp^{kS}}{M}D_{-3dT}^{\perp 2} +S_{LL}D_{-3dLL}^\perp 
  +\frac{k_\perp \cdot S_{LT}}{M}D_{-3dLT}^{\perp2}+\frac{S_{TT}^{kk}}{M^2}D_{-3dTT}^{\perp2}\Big)  \nonumber\\
  &\phantom{XXXXXX} -ic^q_3\Big(\lambda_h D_{+3dL}^\perp +\frac{k_\perp\cdot S_T}{M}D_{+3dT}^{\perp 4}
    +\frac{\varepsilon_\perp^{kS_{LT}}}{M}D_{+3dLT}^{\perp4}+\frac{S_{TT}^{\tilde k k}}{M^2}D_{+3dTT}^{\perp4} \Big)\Big]\nonumber\\
  +&\frac{1}{z(p\cdot q)} k_{\perp\{\mu}\tilde k_{\perp \nu \}} \Big[
      c^q_1 \Big(\lambda_h D_{+3dL}^\perp+\frac{k_\perp\cdot S_T}{M}D_{+3dT}^{\perp 4} +\frac{\varepsilon_\perp^{kS_{LT}}}{M}D_{+3dLT}^{\perp4}
      +\frac{S_{TT}^{\tilde k k}}{M^2}D_{+3dTT}^{\perp4} \Big)  \nonumber\\
  &\phantom{XXXXXX} +ic^q_3\Big(D_{-3d}^\perp+ \frac{\varepsilon_\perp^{kS}}{M}D_{-3dT}^{\perp 2} +S_{LL}D_{-3dLL}^\perp
   +\frac{k_\perp \cdot S_{LT}}{M}D_{-3dLT}^{\perp2}+\frac{S_{TT}^{kk}}{M^2}D_{-3dTT}^{\perp2}\Big)\Big].\label{F:w1L}
\end{align}

For $\tilde W^{(2)}_{\mu\nu}$, we have contributions from $\tilde W^{(2,M)}_{\mu\nu}$,
 $\tilde W^{(2,L)}_{\mu\nu}$ and $\tilde W^{(2,R)}_{\mu\nu}=\tilde W^{(2,L)*}_{\nu\mu}$.
For that from $\tilde W^{(2,M)}_{\mu\nu}$, we calculate
\begin{align}
  & \mathrm{Tr}\big[\hat h^{(2)\rho\sigma}_{\mu\nu} \slashed {\bar n}\big] p^+
  =-8c^q_1p_{\mu}p_{\nu}g_{\perp}^{\rho\sigma}-8ic^q_3p_{\mu}p_{\nu}\varepsilon_{\perp}^{\rho\sigma},\label{F:h2me}\\
  & \mathrm{Tr}\big[\hat h^{(2)\rho\sigma}_{\mu\nu} \gamma^5\slashed {\bar n}\big] p^+
  =8c^q_3p_{\mu}p_{\nu}g_{\perp}^{\rho\sigma} + 8ic^q_1p_{\mu}p_{\nu}\varepsilon_{\perp}^{\rho\sigma},\label{F:h2mo}
\end{align}
and the result is given by
\begin{align}
  \tilde W^{(2,M)}_{t4\mu\nu}=-\frac{4M^2}{z(p\cdot q)^2}p_{\mu}p_{\nu}
  & \Big[ c^q_1\Big(D^M_{+3dd} - \frac{\varepsilon_\perp^{kS}}{M} D_{+3ddT}^{ M\perp}
  +S_{LL}D_{+3ddLL}^M+\frac{k_\perp \cdot S_{LT}}{M}D_{+3ddLT}^{M\perp }+\frac{S_{TT}^{kk}}{M^2}D_{+3ddTT}^{M\perp}\Big)  \nonumber\\
  &-c^q_3  \Big(\lambda_h D_{+3ddL}^M-\frac{k_\perp\cdot S_T}{M} D_{+3ddT}^{M\perp3}
  -\frac{\varepsilon_\perp^{kS_{LT}}}{M}D_{+3ddLT}^{M\perp3}-\frac{S_{TT}^{\tilde k k}}{M^2}D_{+3ddTT}^{M\perp3} \Big)\Big].\label{F:w2m}
\end{align}

For $\tilde W^{(2,L)}_{\mu\nu}$, we have
\begin{align}
  &\mathrm{Tr}\big[\hat N^{(2)\rho\sigma}_{\mu\nu}\slashed {\bar n} \big]p^+
  =4(p\cdot q)c^q_1\Big[g_{\perp}^{\rho\sigma}g_{\perp\mu\nu}+g_{\perp[\mu}^{\rho} g_{\perp\nu]}^{\sigma}\Big]
 -4(p\cdot q)ic^q_3\Big[g_{\perp\mu}^{\rho}\varepsilon_{\perp\nu}^{~~\sigma}-g_{\perp\nu}^{\sigma}\varepsilon_{\perp\mu}^{~~\rho} \Big],\label{F:h2Le}\\
  &\mathrm{Tr}\big[\hat N^{(2)\rho\sigma}_{\mu\nu}\gamma^5\slashed {\bar n} \big]p^+
  =- 4(p\cdot q)c^q_3\Big[g_{\perp}^{\rho\sigma}g_{\perp\mu\nu}+g_{\perp[\mu}^{\rho} g_{\perp\nu]}^{\sigma}\Big]
  +4(p\cdot q)ic^q_1\Big[g_{\perp\mu}^{\rho}\varepsilon_{\perp\nu}^{~~\sigma}-g_{\perp\nu}^{\sigma}\varepsilon_{\perp\mu}^{~~\rho} \Big],\label{F:h2Lo}
\end{align}
and the result is

\begin{align}
  \tilde W^{(2,L)}_{t4\mu\nu} =\frac{2M^2}{z(p\cdot q)} & \bigg\{ g_{\perp\mu\nu} \Big[ c^q_1 \Big(D_{-3dd} - \frac{\varepsilon_\perp^{kS} }{M} D_{-3ddT}^{\perp} +S_{LL}D_{-3ddLL}
  +\frac{k_\perp \cdot S_{LT}}{M}D_{-3ddLT}^{\perp}+\frac{S_{TT}^{kk}}{M^2}D_{-3ddTT}^{\perp}\Big)   \nonumber\\
  &\phantom{XXX}+c^q_3 \Big(\lambda_h D_{-3ddL} -\frac{k_\perp\cdot S_T}{M} D_{-3ddT}^{\perp 3}
  -\frac{\varepsilon_\perp^{kS_{LT}}}{M}D_{-3ddLT}^{\perp3}-\frac{S_{TT}^{\tilde k k}}{M^2}D_{-3ddTT}^{\perp3} \Big)\Big]  \nonumber\\
  + & i\varepsilon_{\perp\mu\nu} \Big[ c^q_1 \Big(\lambda_h D_{-3ddL}-\frac{k_\perp\cdot S_T}{M} D_{-3ddT}^{\perp 3}-\frac{\varepsilon_\perp^{kS_{LT}}}{M}D_{-3ddLT}^{\perp3}
  -\frac{S_{TT}^{\tilde k k}}{M^2}D_{-3ddTT}^{\perp3} \Big)  \nonumber\\
  &\phantom{XXX}+c^q_3\Big(D_{-3dd} - \frac{\varepsilon_\perp^{kS} }{M} D_{-3ddT}^{\perp}
  +S_{LL}D_{-3ddLL}  +\frac{k_\perp \cdot S_{LT}}{M}D_{-3ddLT}^{\perp}+\frac{S_{TT}^{kk}}{M^2}D_{-3ddTT}^{\perp}\Big)\Big]\bigg\}.\label{F:w2L}
\end{align}

We add all the contributions from $\tilde W^{(0)}_{\mu\nu}$, $\tilde W^{(1)}_{\mu\nu}$, and $\tilde W^{(2)}_{\mu\nu}$ together
and use the relationships given by Eqs.~(\ref{eq:t4eomD})-(\ref{eq:t4eomGt}) to eliminate the not independent FFs.
We obtain the twist-4 contributions to the hadronic tensor as given by
\begin{align}
  W_{t4\mu\nu}&=\frac{4M^2}{z(p\cdot q)}\bigg\{   \frac{(zq-2p)_\mu(zq-2p)_\nu}{z^2(p\cdot q)}
  \Big[ c^q_1 \Big(D_3 - \frac{\varepsilon_\perp^{kS}}{M}D^\perp_{3T}
   + S_{LL}D_{3LL}+\frac{k_\perp \cdot S_{LT}}{M}D_{3LT}^{\perp}+\frac{S_{TT}^{kk}}{M^2}D_{3TT}^{\perp} \Big)\nonumber\\
  &\hspace{2cm} -c^q_3\Big(\lambda_hG_{3L} - \frac{k_\perp\cdot S_T}{M}G^\perp_{3T}
  +\frac{\varepsilon_\perp^{kS_{LT}}}{M}G_{3LT}^{\perp}+\frac{S_{TT}^{\tilde k k}}{M^2}G_{3TT}^{\perp} \Big)\Big]\nonumber\\
  &+\frac{k_{\perp\langle\mu}k_{\perp\nu\rangle}}{M^2}
  \Big[ c^q_1\mathrm{Re}\Big(D_{-3d}^\perp+ \frac{\varepsilon_\perp^{kS}}{M}D_{-3dT}^{\perp 2}+S_{LL}D_{-3dLL}^\perp
  +\frac{k_\perp \cdot S_{LT}}{M}D_{-3dLT}^{\perp2}+\frac{S_{TT}^{kk}}{M^2}D_{-3dTT}^{\perp2}\Big) \nonumber\\
  &\hspace{2cm} + c^q_3 \mathrm{Im}\Big(\lambda_h D_{+3dL}^\perp+\frac{k_\perp\cdot S_T}{M}D_{+3dT}^{\perp 4}
  +\frac{\varepsilon_\perp^{kS_{LT}}}{M}D_{+3dLT}^{\perp4}+\frac{S_{TT}^{\tilde k k}}{M^2}D_{+3dTT}^{\perp4}\Big) \Big] \nonumber\\
  &+\frac{k_{\perp\{\mu} \tilde k_{\perp \nu \}}}{2M^2}
  \Big[ c^q_1 \mathrm{Re}\Big(\lambda_h D_{+3dL}^\perp+\frac{k_\perp\cdot S_T}{M}D_{+3dT}^{\perp 4}
  +\frac{\varepsilon_\perp^{kS_{LT}}}{M}D_{+3dLT}^{\perp4}+\frac{S_{TT}^{\tilde k k}}{M^2}D_{+3dTT}^{\perp4}\Big) \nonumber\\
  &\hspace{2cm} -c^q_3 \mathrm{Im}\Big(D_{-3d}^\perp+ \frac{\varepsilon_\perp^{kS}}{M}D_{-3dT}^{\perp 2}
  +S_{LL}D_{-3dLL}^\perp+\frac{k_\perp \cdot S_{LT}}{M}D_{-3dLT}^{\perp2}+\frac{S_{TT}^{kk}}{M^2}D_{-3dTT}^{\perp2}\Big) \Big] \nonumber\\
  &+(c^q_1 g_{\perp\mu\nu}+ic^q_3 \varepsilon_{\perp\mu\nu})
  \mathrm{Re}\Big(D_{-3dd} - \frac{\varepsilon_\perp^{kS} }{M} D_{-3ddT}^{\perp}+S_{LL}D_{-3ddLL}
  +\frac{k_\perp \cdot S_{LT}}{M}D_{-3ddLT}^{\perp}+\frac{S_{TT}^{kk}}{M^2}D_{-3ddTT}^{\perp}\Big)\nonumber\\
  &+(c^q_3 g_{\perp\mu\nu}+ic^q_1 \varepsilon_{\perp\mu\nu})
  \mathrm{Re}\Big(\lambda_h D_{-3ddL}-\frac{k_\perp\cdot S_T}{M} D_{-3ddT}^{\perp 3}
  -\frac{\varepsilon_\perp^{kS_{LT}}}{M}D_{-3ddLT}^{\perp3}-\frac{S_{TT}^{\tilde k k}}{M^2}D_{-3ddTT}^{\perp3}\Big)\bigg\}.\label{eq:t4wsum}
\end{align}
Here a summation over quark flavor is implicit. We can easily check that current conservation $q^\mu W_{t4\mu\nu} = q^\nu W_{t4\mu\nu} = 0$ is valid.

\end{widetext}

\subsection{The structure functions}

By making Lorentz contraction of the hadronic tensor given by Eq.~(\ref{eq:t4wsum}) with the leptonic tensor,
we obtain the differential cross section.
By comparing the results obtained with the general form given by
Eqs.~(\ref{eq:cs})-(\ref{eq:tWTT}) in terms of the structure functions,
we obtain the twist-4 results of the structure functions in QCD parton model at leading order pQCD .
We now present the results. For comparison, we include also the leading twist and twist-3 results here.

Up to twist-4, we have contributions to all the 81 structure functions.
Among them, 18 have both leading twist and twist-4 contributions.
They are given by
\begin{align}
&zW_{U1}=c^e_1 c^q_1 (D_1 -4\kappa_M^2 \mathrm{Re}D_{-3dd} /z ), \label{eq:WU1} \\
  &zW_{U3}=2 c^e_3 c^q_3 (D_1-4\kappa_M^2  \mathrm{Re}D_{-3dd} /z ) ,\label{eq:WU3}\\
  &z\tilde W_{L1}=c^e_1 c^q_3 (G_{1L}-4\kappa_M^2 \mathrm{Re}D_{-3ddL} /z ),\label{eq:tWL1}\\
  &z\tilde W_{L3}=2c^e_3 c^q_1 ( G_{1L}-4\kappa_M^2  \mathrm{Re}D_{-3ddL} /z ),\label{eq:tWL3}\\
  &zW_{T1}^{\sin(\varphi-\varphi_S)}=k_{\perp M}  c^e_1 c^q_1 ( D_{1T}^\perp-4\kappa_M^2 \mathrm{Re}D_{-3ddT}^\perp /z ),\label{eq:WT1}\\
  &zW_{T3}^{\sin(\varphi-\varphi_S)}=2 k_{\perp M} c^e_3 c^q_3 ( D_{1T}^\perp- 4\kappa_M^2  \mathrm{Re}D_{-3ddT}^\perp /z ),\label{eq:WT3}\\
  &z\tilde W_{T1}^{\cos(\varphi-\varphi_S)}=k_{\perp M} c^e_1 c^q_3  (G_{1T}^\perp-4\kappa_M^2 \mathrm{Re}D_{-3ddT}^{\perp 3} /z ),\label{eq:tWT1}\\
  &z\tilde W_{T3}^{\cos(\varphi-\varphi_S)}=2 k_{\perp M} c^e_3 c^q_1  (G_{1T}^\perp-4\kappa_M^2  \mathrm{Re}D_{-3ddT}^{\perp 3} /z ),\label{eq:tWT3}\\
  &zW_{LL1}=c^e_1 c^q_1(D_{1LL}-4\kappa_M^2 \mathrm{Re}D_{-3ddLL} /z ),\label{eq:WLL1}\\
  &zW_{LL3}=2 c^e_3 c^q_3 (D_{1LL}-4\kappa_M^2 \mathrm{Re}D_{-3ddLL} /z ),\label{eq:WLL3}\\
  &zW_{LT1}^{\cos(\varphi-\varphi_{LT})}=-k_{\perp M} c^e_1 c^q_1 (D_{1LT}^\perp-4\kappa_M^2 \mathrm{Re}D_{-3ddLT}^{\perp} /z ),\label{eq:WLT1}\\
  &zW_{LT3}^{\cos(\varphi-\varphi_{LT})}=-2k_{\perp M} c^e_3 c^q_3 (D_{1LT}^\perp-4\kappa_M^2\mathrm{Re}D_{-3ddLT}^{\perp} /z ),\label{eq:WLT3}\\
  &z\tilde W_{LT1}^{\sin(\varphi-\varphi_{LT})}=k_{\perp M} c^e_1 c^q_3 (G_{1LT}^\perp-4\kappa_M^2 \mathrm{Re}D_{-3ddLT}^{\perp3} /z ),\label{eq:tWLT1}\\
  &z\tilde W_{LT3}^{\sin(\varphi-\varphi_{LT})}=2k_{\perp M} c^e_3 c^q_1 (G_{1LT}^\perp-4\kappa_M^2 \mathrm{Re}D_{-3ddLT}^{\perp3} /z ),\label{eq:tWLT3}\\
  &zW_{TT1}^{\cos(2\varphi-2\varphi_{TT})}=k_{\perp M}^2 c^e_1 c^q_1 (D_{1TT}^\perp - 4\kappa_M^2 \mathrm{Re}D_{-3ddTT}^{\perp} /z ),\label{eq:WTT1}\\
  &zW_{TT3}^{\cos(2\varphi-2\varphi_{TT})}=2k_{\perp M}^2 c^e_3 c^q_3 (D_{1TT}^\perp - 4\kappa_M^2 \mathrm{Re}D_{-3ddTT}^{\perp} /z ),\nonumber\\
  &z\tilde W_{TT1}^{\sin(2\varphi-2\varphi_{TT})}=-k_{\perp M}^2 c^e_1 c^q_3 (G_{1TT}^\perp-4\kappa_M^2 \mathrm{Re}D_{-3ddTT}^{\perp3} /z ),\nonumber\\
  &z\tilde W_{TT3}^{\sin(2\varphi-2\varphi_{TT})}=-2k_{\perp M}^2 c^e_3 c^q_1(G_{1TT}^\perp-4\kappa_M^2 \mathrm{Re}D_{-3ddTT}^{\perp3} /z ).\label{eq:tWTT3}
\end{align}
Here, as in \cite{Wei:2016far}, we use $\kappa_M\equiv M/Q$ to symbolize higher twist contributions, i.e., $\kappa_M$ symbolizes twist-3 and $\kappa_M^2$ is twist-4.
We also use $k_{\perp M}\equiv |\vec k_\perp|/M$ to make the equations look more concise.
We may not the quite unfied form of the results obtained for these structure functions. 
Another 27 have only twist-4 contributions
\begin{align}
  &z^2W_{U2}=8 \kappa_M^2  c^e_1 c^q_1 D_3/z, \label{eq:WU2}\\
  &z^2W_{U}^{\cos2\varphi}=-2 k_{\perp M}^2 \kappa_M^2 c^e_1 c^q_1 \mathrm{Re}D_{-3d}^\perp, \label{eq:WUcos2phi}\\
  &z^2\tilde W_{U}^{\sin2\varphi}=-2k_{\perp M}^2 \kappa_M^2  c^e_1 c^q_3 \mathrm{Im}D_{-3d}^\perp, \label{eq:WUsin2phi}\\
  &z^2\tilde W_{L2}=8\kappa_M^2 c^e_1 c^q_3 G_{3L}/z, \label{eq:tWL2}\\
  &z^2\tilde W_L^{\cos2\varphi}=-2  k_{\perp M}^2 \kappa_M^2  c^e_1 c^q_3 \mathrm{Im}D_{+3dL}^\perp,\\
  &z^2W_L^{\sin2\varphi}=2k_{\perp M}^2 \kappa_M^2 c^e_1 c^q_1 \mathrm{Re}D_{+3dL}^\perp,\\
  &z^2W_{T2}^{\sin(\varphi-\varphi_S)}=8 k_{\perp M} \kappa_M^2 c^e_1 c^q_1 D_{3T}^\perp/z,\\
  &z^2W_{T}^{\sin(\varphi+\varphi_S)}=-  k_{\perp M}^3 \kappa_M^2  c^e_1 c^q_1 \mathrm{Re}\big(D_{-3dT}^{\perp 2}+D_{+3dT}^{\perp 4}\big),\\
  &z^2 W_{T}^{\sin(3\varphi-\varphi_S)}= k_{\perp M}^3 \kappa_M^2  c^e_1 c^q_1 \mathrm{Re}\big(D_{-3dT}^{\perp 2}-D_{+3dT}^{\perp 4}\big),\\
  &z^2 \tilde W_{T2}^{\cos(\varphi-\varphi_S)}=8k_{\perp M} \kappa_M^2 c^e_1 c^q_3 G_{3T}^\perp/z,\\
  &z^2\tilde W_{T}^{\cos(\varphi+\varphi_S)}=k_{\perp M}^3 \kappa_M^2 c^e_1 c^q_3 \mathrm{Im}\big(D_{-3dT}^{\perp 2}+D_{+3dT}^{\perp 4}\big),\\
  &z^2\tilde W_{T}^{\cos(3\varphi-\varphi_S)}=-k_{\perp M}^3 \kappa_M^2 c^e_1 c^q_3 \mathrm{Im}\big(D_{-3dT}^{\perp 2}-D_{+3dT}^{\perp 4}\big),\\
  &z^2W_{LL2}=8\kappa_M^2  c^e_1 c^q_1D_{3LL}/z,\\
  &z^2W_{LL}^{\cos2\varphi}=-2k_{\perp M}^2 \kappa_M^2 c^e_1 c^q_1 \mathrm{Re}D_{-3dLL}^\perp,\\
  &z^2\tilde W_{LL}^{\sin2\varphi}=-2k_{\perp M}^2 \kappa_M^2  c^e_1 c^q_3 \mathrm{Im}D_{-3dLL}^\perp,\\
  & z^2W_{LT2}^{\cos(\varphi-\varphi_{LT})}=-8k_{\perp M} \kappa_M^2 c^e_1 c^q_1 D_{3LT}^\perp/z,\\
  & z^2W_{LT}^{\cos(\varphi+\varphi_{LT})}=k_{\perp M}^3 \kappa_M^2 c^e_1 c^q_1 \mathrm{Re}\big(D_{-3dLT}^{\perp 2}-D_{+3dLT}^{\perp 4}\big),\\
  & z^2W_{LT}^{\cos(3\varphi-\varphi_{LT})}=k_{\perp M}^3 \kappa_M^2 c^e_1 c^q_1 \mathrm{Re}\big(D_{-3dLT}^{\perp 2}+D_{+3dLT}^{\perp 4}\big),\\
  &z^2\tilde W_{LT2}^{\sin(\varphi-\varphi_{LT})}=-8k_{\perp M} \kappa_M^2 c^e_1 c^q_3 G_{3LT}^\perp/z,\\
  &z^2\tilde W_{LT}^{\sin(\varphi+\varphi_{LT})}=k_{\perp M}^3 \kappa_M^2 c^e_1 c^q_3 \mathrm{Im}\big(D_{-3dLT}^{\perp 2}-D_{+3dLT}^{\perp 4}\big),\\
  &z^2\tilde W_{LT}^{\sin(3\varphi-\varphi_{LT})}=k_{\perp M}^3 \kappa_M^2 c^e_1 c^q_3 \mathrm{Im}\big(D_{-3dLT}^{\perp 2}+D_{+3dLT}^{\perp 4}\big),\\
  & z^2W_{TT2}^{\cos(2\varphi-2\varphi_{TT})}=8k_{\perp M}^2 \kappa_M^2 c^e_1 c^q_1 D_{3TT}^\perp/z,\\
  & z^2W_{TT}^{\cos2\varphi_{TT}}=-k_{\perp M}^4 \kappa_M^2  c^e_1 c^q_1 \mathrm{Re}\big(D_{-3dTT}^{\perp 2}-D_{+3dTT}^{\perp 4}\big),\\
  &z^2W_{TT}^{\cos(4\varphi-2\varphi_{TT})}=-k_{\perp M}^4 \kappa_M^2 c^e_1 c^q_1 \mathrm{Re}\big(D_{-3dTT}^{\perp 2}+D_{+3dTT}^{\perp 4}\big),\\
  &z^2\tilde W_{TT2}^{\sin(2\varphi-2\varphi_{TT})}=8k_{\perp M}^2 \kappa_M^2 c^e_1 c^q_3 G_{3TT}^\perp/z,\\
  &z^2\tilde W_{TT}^{\sin2\varphi_{TT}}=-k_{\perp M}^4 \kappa_M^2 c^e_1 c^q_3 \mathrm{Im}\big(D_{-3dTT}^{\perp 2}-D_{+3dTT}^{\perp 4}\big), \\
  &z^2\tilde W_{TT}^{\sin(4\varphi-2\varphi_{TT})}=-k_{\perp M}^4 \kappa_M^2 c^e_1 c^q_3 \mathrm{Im}\big(D_{-3dTT}^{\perp 2}+D_{+3dTT}^{\perp 4}\big). \label{eq:tWTTsin4phi-2phiTT}
\end{align}
The rest 36 have only twist-3 contributions
\begin{align}
  &z^2W_{U1}^{\cos\varphi}=4 k_{\perp M} \kappa_M  c^e_3 c^q_3  D^\perp,\label{eq:WU1cosphi}\\
  &z^2W_{U2}^{\cos\varphi}=2k_{\perp M} \kappa_M c^e_1 c^q_1 D^\perp,\\
  &z^2\tilde W_{U1}^{\sin\varphi}=-4k_{\perp M} \kappa_M c^e_3 c^q_1 G^\perp,\\
  &z^2\tilde W_{U2}^{\sin\varphi}=-2k_{\perp M} \kappa_M c^e_1 c^q_3 G^\perp,\\
  &z^2W_{L1}^{\sin\varphi}=4k_{\perp M} \kappa_M c^e_3 c^q_3  D_L^\perp,\\
  &z^2W_{L2}^{\sin\varphi}=2k_{\perp M} \kappa_M c^e_1 c^q_1 D_L^\perp,\\
  &z^2\tilde W_{L1}^{\cos\varphi}=4k_{\perp M} \kappa_M c^e_3 c^q_1 G_L^\perp,\\
  &z^2\tilde W_{L2}^{\cos\varphi}=2k_{\perp M} \kappa_M c^e_1 c^q_3 G_L^\perp,\\
  &z^2W_{T1}^{\sin\varphi_S}=4\kappa_M c^e_3 c^q_3  D_T,\\
  &z^2W_{T2}^{\sin\varphi_S}=2\kappa_M c^e_1 c^q_1  D_T,\\
  &z^2W_{T1}^{\sin(2\varphi-\varphi_S)}=2 k_{\perp M}^2 \kappa_M c^e_3 c^q_3   D_T^\perp,\\
  &z^2W_{T2}^{\sin(2\varphi-\varphi_S)}=k_{\perp M}^2 \kappa_M c^e_1 c^q_1  D_T^\perp,\\
  &z^2\tilde W_{T1}^{\cos\varphi_S}=4\kappa_M c^e_3 c^q_1  G_T,\\
  &z^2\tilde W_{T2}^{\cos\varphi_S}=2\kappa_M c^e_1 c^q_3  G_T,\\
  &z^2\tilde W_{T1}^{\cos(2\varphi-\varphi_S)}=2k_{\perp M}^2 \kappa_M c^e_3 c^q_1 G_T^\perp,\\
  &z^2\tilde W_{T2}^{\cos(2\varphi-\varphi_S)}=k_{\perp M}^2 \kappa_M c^e_1 c^q_3  G_T^\perp,\\
  &z^2W_{LL1}^{\cos\varphi}=4k_{\perp M} \kappa_M c^e_3 c^q_3  D_{LL}^\perp,\\
  &z^2W_{LL2}^{\cos\varphi}=2k_{\perp M} \kappa_M c^e_1 c^q_1  D_{LL}^\perp,\\
  &z^2\tilde W_{LL1}^{\sin\varphi}=-4k_{\perp M} \kappa_M c^e_3 c^q_1  G_{LL}^\perp,\\
  &z^2\tilde W_{LL2}^{\sin\varphi}=-2k_{\perp M} \kappa_M c^e_1 c^q_3  G_{LL}^\perp,\\
  &z^2\tilde W_{LT1}^{\sin\varphi_{LT}}=4\kappa_M c^e_3 c^q_1 G_{LT},\\
  &z^2\tilde W_{LT2}^{\sin\varphi_{LT}}=2\kappa_Mc^e_1 c^q_3 G_{LT},\\
  &z^2\tilde W_{LT1}^{\sin(2\varphi-\varphi_{LT})}=2k_{\perp M}^2 \kappa_M c^e_3 c^q_1 G_{LT}^\perp,\\
  &z^2\tilde W_{LT2}^{\sin(2\varphi-\varphi_{LT})}=k_{\perp M}^2 \kappa_M c^e_1 c^q_3 G_{LT}^\perp,\\
  &z^2W_{LT1}^{\cos\varphi_{LT}}=4\kappa_M c^e_3 c^q_3 D_{LT},\\
  &z^2W_{LT2}^{\cos\varphi_{LT}}=2\kappa_M c^e_1 c^q_1 D_{LT},\\
  &z^2W_{LT1}^{\cos(2\varphi-\varphi_{LT})}=2k_{\perp M}^2 \kappa_M c^e_3 c^q_3 D_{LT}^\perp,\\
  &z^2W_{LT2}^{\cos(2\varphi-\varphi_{LT})}=k_{\perp M}^2 \kappa_M c^e_1 c^q_1 D_{LT}^\perp,\\
  &z^2\tilde W_{TT1}^{\sin(\varphi-2\varphi_{TT})}=4k_{\perp M} \kappa_M c^e_3 c^q_1 G_{TT}^{\prime\perp},\\
  &z^2\tilde W_{TT2}^{\sin(\varphi-2\varphi_{TT})}=2k_{\perp M} \kappa_M c^e_1 c^q_3 G_{TT}^{\prime\perp},\\
  &z^2\tilde W_{TT1}^{\sin(3\varphi-2\varphi_{TT})}=2k_{\perp M}^3 \kappa_M  c^e_3 c^q_1 G_{TT}^\perp,\\
  &z^2\tilde W_{TT2}^{\sin(3\varphi-2\varphi_{TT})}=k_{\perp M}^3 \kappa_M c^e_1 c^q_3 G_{TT}^\perp,\\
  &z^2W_{TT1}^{\cos(\varphi-2\varphi_{TT})}=4k_{\perp M} \kappa_M c^e_3 c^q_3 D_{TT}^{\prime\perp},\\
  &z^2W_{TT2}^{\cos(\varphi-2\varphi_{TT})}=2k_{\perp M} \kappa_M c^e_1 c^q_1 D_{TT}^{\prime\perp}, \\
  &z^2W_{TT1}^{\cos(3\varphi-2\varphi_{TT})}=2k_{\perp M}^3 \kappa_M c^e_3 c^q_3 D_{TT}^\perp, \\
  &z^2W_{TT2}^{\cos(3\varphi-2\varphi_{TT})}=k_{\perp M}^3 \kappa_M c^e_1 c^q_1 D_{TT}^\perp. \label{eq:WTT2}
\end{align}

As in \cite{Wei:2016far} for SIDIS, we see again the following two distinct features:
(1) Structure functions for sine or cosine of even number of azimuthal angles ($\varphi$, $\varphi_S$,  $\varphi_{LT}$ and/or $2\varphi_{TT}$) have leading-twist and/or twist-4 contributions
while those for sine or cosine of odd number of azimuthal angles have twist-3 contributions.
(2) For the structure functions that have leading twist contributions, there are always twist-4 addenda to them. The leading twist and twist-4 contributions mix up with each other.
However, the twist-3 contributions are always separated from the leading twist and/or twist-4 contributions and all of the twist-3 FFs are corresponding to the azimuthal asymmetries which are absent in leading twist and twist-4 contributions.

\subsection{Azimuthal Asymmetries}

Consider the unpolarized case, i.e., summing over the spin of the produced hadron, we have only two twist-3 and two twist-4 azimuthal asymmetries
for $e^+e^-\to h\bar q X$, i.e.,
\begin{align}
  &\langle \cos\varphi \rangle_U = -2k_{\perp M} \kappa_M\frac{ D(y)T^q_2(y) D^\perp}{T^q_0(y) z D_1}, \label{eq:Acosphi}\\
  &\langle \sin\varphi \rangle_U = -2k_{\perp M} \kappa_M \frac{ D(y)T^q_3(y) G^\perp}{T^q_0(y) zD_1}, \label{eq:Asinphi}\\
  &\langle \cos2\varphi \rangle_U = -\frac{1}{2} k_{\perp M}^2 \kappa_M^2 \frac{C(y)c^e_1c^q_1 ~\mathrm{Re} D^\perp_{-3d}}{T^q_0(y)zD_1},  \label{eq:Acos2phi}\\
  &\langle \sin2\varphi \rangle_U =-\frac{1}{2} k_{\perp M}^2 \kappa_M^2 \frac{ C(y)c^e_1c^q_3 ~\mathrm{Im} D^\perp_{-3d}}{T^q_0(y)zD_1}, \label{eq:Asin2phi}
\end{align}
where $y=(1+\cos\theta)/2$,
\begin{align}
  T^q_0(y)&=c^e_1c^q_1A(y)-c^e_3c^q_3B(y), \label{eq:tq0}\\
  T^q_2(y)&=c^e_1c^q_1B(y)-c^e_3c^q_3, \label{eq:tq2}\\
  T^q_3(y)&=c^e_3c^q_1-c^e_1c^q_3B(y), \label{eq:tq3}
\end{align}
and
 $A(y)=(1-y)^2+y^2={(1+\cos^2\theta)}/{2}$, 
  $B(y)=1-2y=-\cos\theta$, 
  $C(y)=4y(1-y)=\sin^2\theta$, 
  $D(y)=\sqrt{y(1-y)}={\sin\theta}/{2}$. 
We note that $\langle \cos\varphi \rangle_U$ and $\langle \cos2\varphi \rangle_U $ are parity conserved,
$\langle \sin\varphi \rangle_U$ and $\langle \sin2\varphi \rangle_U$ are parity violated.

 \subsection{Hadron polarizations}

 We present only results averaged over azimuthal angle $\varphi$.
 For the longitudinal components, we have both leading twist and twist-4 contributions.
 They are given by
 \begin{align}
 &\langle\lambda_h\rangle = -\frac{2}{3} \frac{ P_q(y) T^q_0(y) G_{1L}}{T^q_0(y) D_1} \big(1+\alpha_U \kappa_M^2-\alpha_L \kappa_M^2 \big), \label{eq:lambda}\\
 &\langle S_{LL} \rangle = \frac{1}{2} \frac{T^q_0(y) D_{1LL}}{T^q_0(y) D_1} \big(1+\alpha_U \kappa_M^2-\alpha_{LL} \kappa_M^2 \big); \label{eq:SLL}\\
   &\alpha_U = 4 \frac{ zT^q_0(y) \mathrm{Re}D_{-3dd}-C(y)c^e_1c^q_1 D_{3} }{ z^2T^q_0(y) D_1}, \label{eq:alphaU}\\
  &\alpha_L = 4\frac{ zP_q(y)T_0^q(y) \mathrm{Re}D_{-3ddL}+C(y)c^e_1c^q_3 G_{3L}}{z^2P_q(y)T_0^q(y) G_{1L}}, \label{eq:alphaL}\\
  &\alpha_{LL} = 4 \frac{ zT^q_0(y)\mathrm{Re}D_{-3ddLL}-C(y)c^e_1c^q_1 D_{3LL}}{z^2T^q_0(y) D_{1LL}}, \label{eq:alphaLL}
 \end{align}
 where $P_q(y)$ is the longitudinal polarization of $q$ produced in $e^+e^-\to Z\to q\bar q$, $P_q(y)=T_1^q(y)/T_0^q(y)$, 
 $T^q_1(y)=-c^e_1c^q_3A(y)+c^e_3c^q_1B(y)$.
Here, we emphasize in particular that the factor $T_0^q(y)$ in the numerator and that in the denominator in Eqs.~(\ref{eq:lambda})-(\ref{eq:alphaLL})
can {\it not} cancel with each other since a summation over flavor $q$ is implicit in the numerator and in the denominator,  respectively.
This applies also to all the results presented in the following of this paper.

 For the transverse components with respect to the lepton-hadron plane, we have
 \begin{align}
 &\langle S_T^x \rangle = \frac{8}{3} \kappa_M \frac{ D(y)T^q_3(y) G_{T}}{ T^q_0(y) zD_1}, \label{eq:Stx}\\
 &\langle S_T^y \rangle = -\frac{8}{3} \kappa_M \frac{ D(y)T^q_2(y) D_{T}}{ T^q_0(y) zD_1}, \label{eq:Sty}\\
 &\langle S_{LT}^x \rangle = -\frac{8}{3} \kappa_M \frac{D(y)T^q_2(y)D_{LT}}{ T^q_0(y) z D_1}, \label{eq:Sltx}\\
 &\langle S_{LT}^y \rangle = \frac{8}{3} \kappa_M \frac{D(y)T^q_3(y)G_{LT}}{ T^q_0(y) z D_1}, \label{eq:Slty}\\
 &\langle S_{TT}^{xx} \rangle = -\frac{1}{3} k_{\perp M}^4\kappa_M^2 \frac{C(y)c^e_1c^q_1\mathrm{Re}\big(D_{-3dTT}^{\perp2}-D_{+3dTT}^{\perp4}\big)}{ T^q_0(y)z D_1}, \label{eq:Sttxx}\\
 &\langle S_{TT}^{xy} \rangle = -\frac{1}{3}k_{\perp M}^4\kappa_M^2 \frac{C(y)c^e_1c^q_3\mathrm{Im}\big(D_{-3dTT}^{\perp2}-D_{+3dTT}^{\perp4}\big)}{ T^q_0(y)z D_1}. \label{eq:Sttxy}
 \end{align}
We see that $\langle S_T^x \rangle$, $\langle S_T^y \rangle$, $\langle S_{LT}^x \rangle$ and $\langle S_{LT}^y \rangle$ have only twist-3 contributions
while $\langle S_{TT}^{xx} \rangle$ and $\langle S_{TT}^{xy} \rangle$ have only twist-4 contributions.

For the transverse components with respect to the hadron-jet plane, we obtain
\begin{align}
  &\langle S_T^n \rangle =\frac{2}{3} k_{\perp M} \frac{T^q_0(y)D_{1T}^\perp}{T^q_0(y) D_1} \big(1+\alpha_U \kappa_M^2-\alpha_{T}^n \kappa_M^2 \big),\label{eq:Stn} \\
  &\langle S_T^t \rangle =-\frac{2}{3} k_{\perp M} \frac{P_q(y)T_0^q(y)G_{1T}^\perp}{ T^q_0(y) D_1}\big(1+\alpha_U \kappa_M^2-\alpha_{T}^t \kappa_M^2 \big),\label{eq:Stt} \\
  &\langle S_{LT}^n \rangle =-\frac{2}{3} k_{\perp M} \frac{P_q(y)T_0^q(y)G_{1LT}^\perp}{ T^q_0(y) D_1} \big(1+\alpha_U \kappa_M^2-\alpha_{LT}^n \kappa_M^2 \big),\label{eq:Sltn} \\
  &\langle S_{LT}^t \rangle =-\frac{2}{3} k_{\perp M} \frac{T^q_0(y)D_{1LT}^\perp}{ T^q_0(y) D_1} \big(1+\alpha_U \kappa_M^2-\alpha_{LT}^t \kappa_M^2 \big),\label{eq:Sltt} \\
  &\langle S_{TT}^{nn} \rangle =-\frac{2}{3} k_{\perp M}^2 \frac{T^q_0(y)D_{1TT}^\perp}{ T^q_0(y) D_1} \big(1+\alpha_U \kappa_M^2-\alpha_{TT}^{nn} \kappa_M^2 \big),\label{eq:Sttnn} \\
  &\langle S_{TT}^{nt} \rangle =\frac{2}{3} k_{\perp M}^2 \frac{P_q(y)T_0^q(y)G_{1TT}^\perp}{ T^q_0(y) D_1} \big(1+\alpha_U \kappa_M^2-\alpha_{TT}^{nt} \kappa_M^2 \big),\label{eq:Sttnt}
\end{align}
where the $\alpha$'s are similar to those given by Eqs.~(\ref{eq:alphaL})-(\ref{eq:alphaLL}) in the longitudinally polarized case, i.e.,
\begin{align}
  &\alpha_{T}^n =4 \frac{ zT^q_0(y)\mathrm{Re}D_{-3ddT}^\perp-C(y)c^e_1c^q_1~D_{3T}^\perp}{z^2T^q_0(y)D_{1T}^\perp}, \label{eq:alphaTn}\\
  &\alpha_{T}^t = 4\frac{zP_q(y)T_0^q(y)\mathrm{Re}D_{-3ddT}^{\perp 3}+C(y)c^e_1c^q_3~G_{3T}^\perp}{z^2P_q(y)T_0^q(y)G_{1T}^\perp}, \label{eq:alphaTt}\\
  &\alpha_{LT}^n = 4\frac{zP_q(y)T_0^q(y)\mathrm{Re}D_{-3ddLT}^{\perp3}-C(y)c^e_1c^q_3~G_{3LT}^\perp}{z^2P_q(y)T_0^q(y)G_{1LT}^\perp}, \label{eq:alphaLTn}\\
  &\alpha_{LT}^t = 4\frac{zT^q_0(y)\mathrm{Re}D_{-3ddLT}^{\perp}-C(y)c^e_1c^q_1~D_{3LT}^\perp}{z^2T^q_0(y)D_{1LT}^\perp}, \label{eq:alphaLTt}\\
  &\alpha_{TT}^{nn} = 4\frac{zT^q_0(y)\mathrm{Re}D_{-3ddTT}^\perp-C(y)c^e_1c^q_1D_{3TT}^\perp}{ z^2T^q_0(y)D_{1TT}^\perp}, \label{eq:alphaTTn}\\
  &\alpha_{TT}^{nt} = 4\frac{zP_q(y)T_0^q(y)\mathrm{Re}D_{-3ddTT}^{\perp 3}-C(y)c^e_1c^q_3G_{3TT}^\perp}{z^2P_q(y)T_0^q(y)G_{1TT}^\perp}. \label{eq:alphaTTt}
\end{align}
We see that the transverse components with respect to the hadron-jet plane have  both leading and twist-4 contributions.
We also note that the leading twist and twist-3 parts are the same as those obtained in \cite{Wei:2014pma,Chen:2016moq}.

If we use the relationships given by Eqs.~(\ref{eq:g0D3d})-(\ref{eq:g0G3ddST}) obtained at $g=0$, we obtain
\begin{align}
 &\alpha_U\approx -k_{\perp M}^2 \Bigg[\frac{\partial\ln{T^q_0(y)D_1}}{\partial{\ln z}} +  \frac{2C(y)c^e_1c^q_1D_1}{T^q_0(y)D_1}\Bigg], \label{eq:alphaUg=0}\\
 &\alpha_L\approx -k_{\perp M}^2 \Bigg[\frac{\partial\ln{P_q(y)T_0^q(y)G_{1L}}}{\partial{\ln z}} -  \frac{2C(y)c^e_1c^q_3G_{1L}}{P_q(y)T_0^q(y)G_{1L}}\Bigg], \label{eq:alphaLg=0}\\
 &\alpha_{LL}\approx -k_{\perp M}^2 \Bigg[\frac{\partial\ln{T^q_0(y)D_{1LL}}}{\partial{\ln z}} +  \frac{2C(y)c^e_1c^q_1D_{1LL}}{T^q_0(y)D_{1LL}}\Bigg], \label{eq:alphaLLg=0}\\
 &\alpha_T^n\approx -k_{\perp M}^2 \Bigg[\frac{\partial\ln{T^q_0(y)D_{1T}^\perp}}{\partial{\ln z}} +  \frac{2C(y)c^e_1c^q_1D_{1T}^\perp}{T^q_0(y)D_{1T}^\perp}\Bigg], \label{eq:alphaTng=0}\\
 &\alpha_T^t\approx -k_{\perp M}^2 \Bigg[\frac{\partial\ln{P_q(y)T_0^q(y)G_{1T}^\perp}}{\partial{\ln z}} -  \frac{2C(y)c^e_1c^q_3G_{1T}^\perp}{P_q(y)T_0^q(y)G_{1T}^\perp}\Bigg], \label{eq:alphaTtg=0}\\
 &\alpha_{LT}^n\approx -k_{\perp M}^2 \Bigg[\frac{\partial\ln{P_q(y)T_0^q(y)G_{1LT}^\perp}}{\partial{\ln z}} - \frac{2C(y)c^e_1c^q_3G_{1LT}^\perp}{P_q(y)T_0^q(y)G_{1LT}^\perp}\Bigg], \nonumber\\
 &\alpha_{LT}^t\approx -k_{\perp M}^2 \Bigg[\frac{\partial\ln{T^q_0(y)D_{1LT}^\perp}}{\partial{\ln z}} +  \frac{2C(y)c^e_1c^q_1D_{1LT}^\perp}{T^q_0(y)D_{1LT}^\perp}\Bigg], \label{eq:alphaLTng=0}\\
 &\alpha_{TT}^{nt}\approx -k_{\perp M}^2 \Bigg[\frac{\partial\ln{P_q(y)T_0^q(y)G_{1TT}^\perp}}{\partial{\ln z}} - \frac{2C(y)c^e_1c^q_3G_{1TT}^\perp}{P_q(y)T_0^q(y)G_{1TT}^\perp}\Bigg], \nonumber\\\label{eq:alphaTTntg=0}
 &\alpha_{TT}^{nn}\approx -k_{\perp M}^2 \Bigg[\frac{\partial\ln{T^q_0(y)D_{1TT}^\perp}}{\partial{\ln z}} + \frac{2C(y)c^e_1c^q_1D_{1TT}^\perp}{T^q_0(y)D_{1TT}^\perp}\Bigg].
\end{align}

\begin{figure}
  \centering
   \includegraphics[width=5cm]{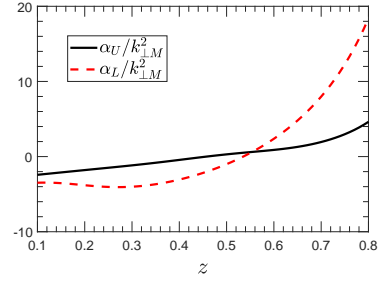}\\
  \caption{A rough estimation of the twist-4 contribution factor $\alpha/k_{\perp M}^2$ as a function of $z$ at $y=0.5$ and $Q=M_Z$.} \label{alphaU}
\end{figure}

At present stage, we may use these equations to make rough estimations for twist-4 contributions.
To get a feeling of how large they could be, we plot $\alpha_U$ and $\alpha_L$ using the parameterizations of FFs in \cite{deFlorian:1997zj,Albino:2008fy,Chen:2016iey}.
We see from Fig.~\ref{alphaU} that the modifications could be quite significant.

\subsection{Contributions from the four-quark correlator}\label{sec:4q}

\begin{figure}
  \centering
  \includegraphics[width=0.45\textwidth]{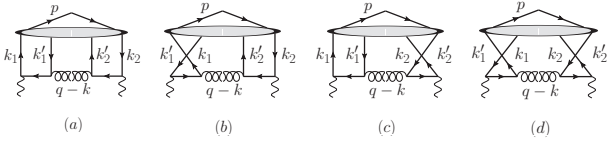}
  \caption{The first four of the four-quark diagrams where no multiple gluon scattering is involved.
  In (a), we have $k_1'=k_1-k$ and $k_2'=k_2-k$;
  in (b)  we have the interchange of $k_1$ with $k_1'$;
  in (c) we have the interchange of $k_2$ with $k_2'$;
  in (d) we have both interchanges of $k_1$ with $k_1'$ and $k_2$ with $k_2'$. }\label{fig:4q}
\end{figure}

The calculations presented above are made only for $e^+e^-\to h\bar q X$ where only quark-$j$-gluon-quark correlators are considered.
Similar to those in deeply inelastic lepton-nucleon scattering discussed in \cite{Qiu:1988dn} and \cite{Wei:2016far},
up to twist-4, we have also contributions from diagrams involving the four quark correlator
\begin{align}
  \hat \Xi^{(0)}_{(4q)}&(k_1,k,k_2)=\frac{g^2}{8}\int\frac{d^4y}{(2\pi)^4}\frac{d^4y_1}{(2\pi)^4}\frac{d^4y_2}{(2\pi)^4} e^{-ik_1y+i(k_1-k)y_1-i(k_2-k)y_2} \nonumber\\
  &\sum_X\langle 0|\bar\psi(y_2)\mathcal{L}^\dag(0,y_2)\psi(0)|hX \rangle\langle hX|\bar\psi(y)\mathcal{L}(y,y_1)\psi(y_1)|0 \rangle. \label{eq:4qcorr}
\end{align}
Example of such diagrams are shown in Fig.~\ref{fig:4q}
where we obtain contributions to $e^+e^-\to hg X$ if the cut is given at the middle
while they contribute to $e^+e^-\to h\bar q X$ if we have the left or right cut.
Both of them contribute to $e^+e^-\to h+\mathrm{jet}+X$, so we consider them together.

It can be shown that the collinear expansion can also be applied to this case and the gauge links included
in the correlators given by Eq.~(\ref{eq:4qcorr}) are obtained by taking the multiple gluon scattering  into account.
The hadronic tensor $W_{4q\mu\nu}^{(g)}$ for $e^+e^-\to h+g+X$ and $W_{4q\mu\nu}^{(q)}$ for $e^+e^-\to h+\bar q+X$
can be written as the unified form
\begin{align}
  W_{4q\mu\nu}^{(g/q)}=\frac{1}{p\cdot q} &\int dzdz_1dz_2  h^{g/q}_{4q}\Big[\big(c_1^q g_{\perp\mu\nu}+i c_3^q \varepsilon_{\perp\mu\nu} \big)C_s\nonumber\\
  +& \big(c_3^q g_{\perp\mu\nu}+ic_1^q \varepsilon_{\perp\mu\nu}\big)C_{ps}\Big]. \label{F:W4q-uni}
\end{align}
Here $C_s$ and $C_{ps}$ are TMD correlation functions given by
\begin{align}
  C_j=&\int d^4k_1d^4kd^4k_2\delta(z-\frac{p^+}{k^+})\delta(k_1^+ z_1-p^+)\delta(k_2^+ z_2-p^+)\nonumber\\
  &\times(2\pi)^2\delta^2(\vec k_\perp + \vec k_\perp')\Xi^{(0)}_{(4q)j}(k_1,k,k_2;p,S), \label{eq:Cs-ps}
\end{align}
where $j=s$ or $ps$ and the unintegrated correlation functions  $\Xi^{(0)}_{(4q)s}$ and $\Xi^{(0)}_{(4q)ps}$ are defined as
\begin{align}
  \Xi^{(0)}_{(4q)s}&=\frac{g^2}{8}\int\frac{d^4y}{(2\pi)^4}\frac{d^4y_1}{(2\pi)^4}\frac{d^4y_2}{(2\pi)^4} e^{-ik_1y+i(k_1-k)y_1-i(k_2-k)y_2} \nonumber\\
  &\sum_X \Big\{\langle 0|\bar\psi(y_2)\slashed n\psi(0)|hX \rangle \langle hX|\bar\psi(y)\slashed n\psi(y_1)|0 \rangle \nonumber\\
   &+\langle 0|\bar\psi(y_2)\gamma^5\slashed n\psi(0)|hX \rangle \langle hX|\bar\psi(y)\gamma^5\slashed n\psi(y_1)|0 \rangle  \Big\}, \label{eq:Xis}\\
  \Xi^{(0)}_{(4q)ps}&=\frac{g^2}{8}\int\frac{d^4y}{(2\pi)^4}\frac{d^4y_1}{(2\pi)^4}\frac{d^4y_2}{(2\pi)^4} e^{-ik_1y+i(k_1-k)y_1-i(k_2-k)y_2} \nonumber\\
  &\sum_X \Big\{\langle 0|\bar\psi(y_2)\gamma^5\slashed n\psi(0)|hX \rangle \langle hX|\bar\psi(y)\slashed n\psi(y_1)|0 \rangle \nonumber\\
   &+\langle 0|\bar\psi(y_2)\slashed n\psi(0)|hX \rangle \langle hX|\bar\psi(y)\gamma^5\slashed n\psi(y_1)|0 \rangle  \Big\}, \label{eq:Xips}
\end{align}
where we have omitted the gauge links that are the same as those in Eq.~(\ref{eq:4qcorr}).
The $h^{g/q}_{4q}$ are obtained by summing over all the diagrams. For $h^g_{4q}$, we obtain
\begin{align}
  h_{4q}^{g}&=\frac{z z_B^3 \delta(z-z_B)}{\big(z_1-z_B+i\epsilon\big)\big(z_2-z_B-i\epsilon\big)}+ \frac{z_B^2/z_1 z_2 \delta(z-z_B)}{\big(1/z_1+i\epsilon\big)\big(1/z_2-i\epsilon\big)} \nonumber\\
  &- \frac{z_B^3/z_2\delta(z-z_B)}{(z_1-z_B+i\epsilon)(1/z_2-i\epsilon)}-(1\leftrightarrow 2)^*. \label{eq:h4g}
  \end{align}
For $h^{q}_{4q}$, we have, $h_{4q}^q=h_{4q}^{qL}+h_{4q}^{qR}$,
\begin{align}
  h_{4q}^{qL}&=\frac{z z_B^3 \delta(z_1-z_B)}{\big(z-z_B-i\epsilon\big)\big(z_2-z_B-i\epsilon\big)}- \Big(\frac{1}{z_2}\to \frac{1}{z}-\frac{1}{z_2}\Big) \nonumber\\
  &- \frac{z z_B^3 \delta(z_1 +z_B-\frac{z_1z_B}{z})}{\big(z-z_B-i\epsilon\big)\big(z_2-z_B-i\epsilon\big)} +\Big(\frac{1}{z_2}\to \frac{1}{z}-\frac{1}{z_2}\Big),\label{eq:h4q}
\end{align}
and $h_{4q}^{qR}(z_1,z,z_2)=h_{4q}^{qL*}(z_2,z,z_1)$.
Adding all of them together, we obtain $h_{4q}=h_{4q}^{qL}+h_{4q}^{qR}+h_{4q}^g$. For $C_s$ and $C_{ps}$, they can be decomposed as
\begin{align}
  z\int dzdz_1dz_2 & h_{4q}C_{s}=M^2\Big(D_{4q}-\frac{\varepsilon_\perp^{kS}}{M}D_{4qT}^\perp+S_{LL}D_{4qLL} \nonumber\\
     &+\frac{k_\perp \cdot S_{LT}}{M}D_{4qLT}^\perp +\frac{S_{TT}^{kk}}{M^2}D_{4qTT}^\perp \Big)
     ,\label{eq:cs4q}\\
  z\int dzdz_1dz_2 & h_{4q}C_{ps}=M^2\Big(\lambda_h G_{4qL}-\frac{k_\perp \cdot S_T}{M}G_{4qT}^{\perp} \nonumber\\
     &+\frac{\varepsilon_\perp^{kS_{LT}}}{M}G_{4qLT}^{\perp}+\frac{S_{TT}^{\tilde kk}}{M^2}G_{4qTT}^{\perp } \Big).\label{eq:ps4q}
\end{align}
The contributions to the structure functions are given by
\begin{align}
  &z^2 W_{4qU1}= -\kappa_M^2 c_1^e c_1^q D_{4q}, \label{eq:4q-wu1} \\
  &z^2 W_{4qU3}= -2\kappa_M^2 c_3^e c_3^q D_{4q},\label{eq:4q-wu3}\\
  &z^2 \tilde W_{4qL1}=-\kappa_M^2 c_1^e c_3^q G_{4qL} ,\label{EQ:4q-wV1}\\
  &z^2 \tilde W_{4qL3}= -2\kappa_M^2 c_3^e c_1^q G_{4qL} ,\label{EQ:4q-wV3}\\
  &z^2 W_{4qT1}^{\sin(\varphi-\varphi_S)}=-k_{\perp M}\kappa_M^2 c_1^e c_1^q D_{4qT}^{\perp}, \label{EQ:4q-wT1}\\
  &z^2 W_{4qT3}^{\sin(\varphi-\varphi_S)}=-2 k_{\perp M}\kappa_M^2 c_3^e c_3^q D_{4qT}^\perp, \label{EQ:4q-wT3}\\
  &z^2 \tilde W_{4qT1}^{\cos(\varphi-\varphi_S)}=-k_{\perp M}\kappa_M^2 c_1^e c_3^q G_{4qT}^{\perp},\label{EQ:4q-wtT1}\\
  &z^2 \tilde W_{4qT3}^{\cos(\varphi-\varphi_S)}=-2k_{\perp M}\kappa_M^2 c_3^e c_1^q G_{4qT}^{\perp} ,\label{EQ:4q-wtT3}\\
  &z^2 W_{4qLL1}=-\kappa_M^2 c_1^e c_1^q D_{4qLL}, \label{EQ:4q-wLL1} \\
  &z^2 W_{4qLL3}=-2\kappa_M^2 c_3^e c_3^q D_{4qLL}, \label{EQ:4q-wLL3}\\
  &z^2 \tilde W_{4qLT1}^{\sin(\varphi-\varphi_{LT})}=k_{\perp M}\kappa_M^2 c_1^e c_3^q G_{4qLT}^{\perp}, \label{EQ:4q-twLT1}\\
  &z^2 \tilde W_{4qLT3}^{\sin(\varphi-\varphi_{LT})}= 2 k_{\perp M}\kappa_M^2 c_3^e c_1^q G_{4qLT}^{\perp}, \label{EQ:4q-twLT3}\\
  &z^2 W_{4qLT1}^{\cos(\varphi-\varphi_{LT})}=k_{\perp M}\kappa_M^2 c_1^ec_1^q D_{4qLT}^{\perp}, \label{EQ:4q-wLT1}\\
  &z^2 W_{4qLT3}^{\cos(\varphi-\varphi_{LT})}=2k_{\perp M}\kappa_M^2 c_3^e c_3^q D_{4qLT}^{\perp}, \label{EQ:4q-wLT3}\\
  &z^2 \tilde W_{4qTT1}^{\sin(2\varphi-2\varphi_{TT})}=-k_{\perp M}^2\kappa_M^2 c_1^e c_3^q G_{4qTT}^{\perp},\label{EQ:4q-twTT1}\\
  &z^2 \tilde W_{4qTT3}^{\sin(2\varphi-2\varphi_{TT})}=-2 k_{\perp M}^2\kappa_M^2 c_3^e c_1^q G_{4qTT}^{\perp}, \label{EQ:4q-twTT3}\\
  &z^2 W_{4qTT1}^{\cos(2\varphi-2\varphi_{TT})}=-k_{\perp M}^2\kappa_M^2c_1^ec_1^q D_{4qTT}^{\perp}, \label{eq:4q-wTT1}\\
  &z^2 W_{4qTT3}^{\cos(2\varphi-2\varphi_{TT})}=-2k_{\perp M}^2\kappa_M^2c_3^ec_3^q D_{4qTT}^{\perp}. \label{eq:4q-wTT3}
\end{align}
We see that they have the same modes as for the leading twist contributions.
They lead to twist-4 modifications of hadron polarizations which are given by
\begin{align}
 &\alpha_{4qU}= \frac{T^q_0(y)D_{4q} }{zT^q_0(y)D_1}, &&\nonumber \\
 &\alpha_{4qL}= \frac{P_q(y)T_0^q(y)G_{4qL} }{zP_q(y)T_0^q(y)G_{1L}}, &&\alpha_{4qLL}=\frac{T^q_0(y)D_{4qLL} }{zT^q_0(y)D_{1LL}},  \nonumber \\
 &\alpha_{4qT}^t= \frac{P_q(y)T_0^q(y)G_{4qT}^{\perp} }{zP_q(y)T_0^q(y)G_{1T}^\perp}, &&\alpha_{4qT}^n= \frac{ T^q_0(y)D_{4qT}^{\perp} }{zT^q_0(y)D_{1T}^\perp}, \nonumber\\
 &\alpha_{4qLT}^n=-\frac{P_q(y)T_0^q(y)G_{4qLT}^{\perp} }{zP_q(y)T_0^q(y)G_{1LT}^\perp}, &&\alpha_{4qLT}^t= \frac{T^q_0(y)D_{4qLT}^{\perp} }{zT^q_0(y)D_{1LT}^\perp},\nonumber\\
 &\alpha_{4qTT}^{nt}= \frac{P_q(y)T_0^q(y) G_{4qTT}^{\perp} }{zP_q(y)T_0^q(y)G_{1LT}^\perp},&&\alpha_{4qTT}^{nn}=\frac{ T^q_0(y)D_{4qTT}^{\perp} }{zT^q_0(y)D_{1TT}^\perp}.\nonumber
\end{align}

\subsection{Reducing to the inclusive process}

By integrating the differential cross section for the semi-inclusive process $e^+e^-\to h\bar qX$ over $d^2 k'_\perp$,
we obtain that for the inclusive process  $e^+e^-\to hX$ and correspondingly the inclusive structure functions given by
Eqs.~(\ref{eq:FUin})-(\ref{eq:tFTTin}).
Among the 19 inclusive structure functions, 6 of them have leading twist contributions, they are given by
\begin{align}
  &zF^{(in)}_{U1}=c^e_1 c^q_1\Big[\hat D_1 -\kappa_M^2\big(4\mathrm{Re} \hat D_{-3dd}+ \hat{D}_{4q}\big)/z \Big],\label{eq:SFun1}\\
  &zF^{(in)}_{U3}=2c^e_3 c^q_3\Big[\hat D_1 -\kappa_M^2\big( 4\mathrm{Re} \hat D_{-3dd}+ \hat{D}_{4q}\big)/z \Big],\label{eq:SFun3}\\
  &z\tilde F^{(in)}_{L1}=c^e_1 c^q_3 \Big[\hat G_{1L} -\kappa_M^2\big( 4\mathrm{Re} \hat D_{-3ddL}+ \hat{G}_{4qL}\big)/z \Big] ,\label{eq:SFL1}\\
  &z\tilde F^{(in)}_{L3}=2c^e_3 c^q_1\Big[\hat G_{1L} -\kappa_M^2\big( 4\mathrm{Re} \hat D_{-3ddL}+ \hat{G}_{4qL}\big)/z \Big],  \label{EQ:SFL3}\\
   &zF^{(in)}_{LL1}=c^e_1 c^q_1 \Big[\hat D_{1LL} - \kappa_M^2\big(4 \mathrm{Re} \hat D_{-3ddLL}+ \hat{D}_{4qLL}\big)/z \Big],\label{EQ:SFLL1}\\
  &zF^{(in)}_{LL3}=2c^e_3 c^q_3 \Big[\hat D_{1LL} -\kappa_M^2\big(4\mathrm{Re} \hat D_{-3ddLL}+ \hat{D}_{4qLL}\big)/z \Big],\label{EQ:SFLL3}
  \end{align}
where $\hat D$'s and $\hat G$'s are the corresponding one-dimensional FFs that can be
obtained by integrating their three-dimensional counterparts over $d^2k'_\perp /(2\pi)^2$.
We see that these 6 structure functions correspond to the unpolarized, the longitudinally polarized and $S_{LL}$-dependent case.
In Eqs.~(\ref{eq:FUin})-(\ref{eq:FLLin}) they correspond to the $(1+\cos^2\theta)$ and $\cos\theta$-terms.

There are 8 structure functions have twist-3 contributions, and are given by
\begin{align}
  &z^2F_{T1}^{(in)\sin\varphi_S}=4\kappa_Mc^e_3 c^q_3 \hat D_T,\label{EQ:SFT1}\\
  &z^2F_{T2}^{(in)\sin\varphi_S}=2\kappa_Mc^e_1 c^q_1 \hat D_T,\label{EQ:SFT2}\\
  &z^2\tilde F_{T1}^{(in)\cos\varphi_S}=4\kappa_Mc^e_3 c^q_1 \hat G_T,\label{EQ:SFtT1}\\
  &z^2\tilde F_{T2}^{(in)\cos\varphi_S}=2\kappa_Mc^e_1 c^q_3 \hat G_T,\label{EQ:SFtT2}\\
  &z^2\tilde F_{LT1}^{(in)\sin\varphi_{LT}} = 4\kappa_Mc^e_3c^q_1\hat G_{LT},\label{EQ:SFtLT1}\\
  &z^2\tilde F_{LT2}^{(in)\sin\varphi_{LT}} = 2\kappa_Mc^e_1c^q_3\hat G_{LT},\label{EQ:SFtLT2}\\
  &z^2F_{LT1}^{(in)\cos\varphi_{LT}} = 4\kappa_Mc^e_3c^q_3\hat D_{LT},\label{EQ:SFLT1}\\
  &z^2F_{LT2}^{(in)\cos\varphi_{LT}} = 2\kappa_Mc^e_1c^q_1\hat D_{LT}. \label{EQ:SFLT2}
 \end{align}
They all correspond to the transverse components of hadron polarization,
where 4 of them correspond to the transverse components of the vector polarization
with respect to the hadron-lepton plane and another 4 correspond to the $S_{LT}$-dependent part.
In Eqs.~(\ref{eq:FTin})-(\ref{eq:tFLTin}) they correspond to the $\sin\theta$ and $\sin2\theta$-terms.

The rest 5 of the 19 strcture functions have only twist-4 contributions, and they are given by
\begin{align}
  &zF^{(in)}_{U2}=8\kappa_M^2 c^e_1 c^q_1 \hat D_3/z^2,\label{EQ:SFun2}\\
  &z\tilde F^{(in)}_{L2}=8\kappa_M^2 c^e_1 c^q_3 \hat G_{3L}/z^2,\label{EQ:SFL2}\\
  &zF^{(in)}_{LL2}=8\kappa_M^2 c^e_1 c^q_1 \hat D_{3LL}/z^2,\label{EQ:SFLL2}\\
  &z^2F_{TT}^{(in)\cos2\varphi_{TT}}=-\kappa_M^2  c^e_1 c^q_1 \mathrm{Re}\big(\hat D_{-3dTT}^{\perp 2}-\hat D_{+3dTT}^{\perp 4}\big),\label{eq:WTTcos2phiTTin}\\
  &z^2\tilde F_{TT}^{(in)\sin2\varphi_{TT}}=-\kappa_M^2 c^e_1 c^q_3 \mathrm{Im}\big(\hat D_{-3dTT}^{\perp 2}-\hat D_{+3dTT}^{\perp 4}\big). \label{eq:WTTsin2phiTTin}
\end{align}
They all correspond to the $\sin^2\theta$-terms in Eqs.~(\ref{eq:FUin})-(\ref{eq:FLLin}) and Eqs.~(\ref{eq:FTTin})-(\ref{eq:tFTTin}).

These results show that in $e^+e^-\to Z\to h+X$, we have leading twist longitudinal polarization and spin alignment with twist-4 addenda.
They are given by
\begin{align}
 &\langle\lambda_h\rangle^{(in)} = -\frac{2}{3} \frac{ P_q(y) T^q_0(y) \hat G_{1L}}{T^q_0(y) \hat D_1} \big(1+\alpha_U^{(in)} \kappa_M^2-\alpha_L^{(in)}  \kappa_M^2 \big), \label{eq:lambdainclusive}\\
 &\langle S_{LL} \rangle^{(in)}  = \frac{1}{2} \frac{T^q_0(y) \hat D_{1LL}}{T^q_0(y) \hat D_1} \big(1+\alpha_U^{(in)}  \kappa_M^2-\alpha_{LL}^{(in)}  \kappa_M^2 \big); \label{eq:SLLinclusive}\\
   &\alpha_U^{(in)} =  \frac{ zT^q_0(y) (4\mathrm{Re}\hat D_{-3dd}+\hat D_{4q})-4C(y)c^e_1c^q_1 \hat D_{3} }{ z^2T^q_0(y) \hat D_1}, \label{eq:alphaUinclusive}\\
   &\alpha_L^{(in)}  = \frac{ zP_q(y)T_0^q(y) (4\mathrm{Re}\hat D_{-3ddL}+\hat G_{4qL})+4C(y)c^e_1c^q_3 \hat G_{3L}}{z^2P_q(y)T_0^q(y) \hat G_{1L}}, \label{eq:alphaLinclusive}\\
   &\alpha_{LL}^{(in)}  = \frac{ zT^q_0(y) (4\mathrm{Re}\hat D_{-3ddLL}+\hat D_{4qLL})-4C(y)c^e_1c^q_1 \hat D_{3LL}}{z^2T^q_0(y) \hat D_{1LL}}. \label{eq:alphaLLinclusive}
 \end{align}
We have also twist-3 transverse polarization with respect to the lepton-hadron plane given by
 \begin{align}
 &\langle S_T^x \rangle^{(in)}  = \frac{8}{3} \kappa_M \frac{ D(y)T^q_3(y) \hat G_{T}}{ T^q_0(y) z\hat D_1}, \label{eq:Stxinclusive}\\
 &\langle S_T^y \rangle^{(in)}  = -\frac{8}{3}\kappa_M \frac{ D(y)T^q_2(y) \hat D_{T}}{ T^q_0(y) z\hat D_1}, \label{eq:Styinclusive}\\
 &\langle S_{LT}^x \rangle^{(in)}  = -\frac{8}{3} \kappa_M \frac{D(y)T^q_2(y) \hat D_{LT}} {T^q_0(y) z \hat D_1}, \label{eq:Sltxinclusive}\\
 &\langle S_{LT}^y \rangle^{(in)}  = \frac{8}{3} \kappa_M \frac{D(y)T^q_3(y) \hat G_{LT}} {T^q_0(y) z \hat D_1}. \label{eq:Sltyinclusive}
 \end{align}
 For the $S_{TT}$-components, we have only twist-4 contributions
 \begin{align}
 &\langle S_{TT}^{xx} \rangle^{(in)}  =-\frac{1}{3} \kappa_M^2 \frac{C(y)c^e_1c^q_1\mathrm{Re}\big(\hat D_{-3dTT}^{\perp2}-\hat D_{+3dTT}^{\perp4}\big)}{ T^q_0(y)z \hat D_1}, \label{eq:Sttxxinclusive}\\
 &\langle S_{TT}^{xy} \rangle^{(in)}  =-\frac{1}{3} \kappa_M^2 \frac{C(y)c^e_1c^q_3\mathrm{Im}\big(\hat D_{-3dTT}^{\perp2}-\hat D_{+3dTT}^{\perp4}\big)}{T^q_0(y)z \hat D_1}. \label{eq:Sttxyinclusive}
\end{align}
It is interesting to see that even for the inclusive reaction, we can study the twist-3 and twist-4 FFs by measuring these different components of hadron polarization.

\section{Summary}\label{sec:summary}

 We present the complete twist-4 results for the semi-inclusive annihilation process $e^+ +e^- \rightarrow h+ \bar q(\mathrm{jet})+X$.
 The calculations have been carried out by using the collinear expansion where the multiple gluon scattering
 have been taken into account and gauge links are obtained systematically and automatically.
 We present the cross section in terms of structure functions and the structure functions are given in terms of the gauge invariant FFs.

Among the 81 structure functions, 18 of them have both leading twist and twist-4 contributions; 27 have only twist-4 contributions and
the rest 36 have twist-3 contributions.
All those correspond to sine or cosine of even number of azimuthal angle have leading twist and/or twist-4 contributions;
those correspond to sine or cosine of odd number of azimuthal angle have twist-3 contributions.
For any structure function that has leading twist contribution, there is a twist-4 addendum to it.

We also present the results of azimuthal asymmetries and different components of hadron polarization in terms of gauge invariant FFs.
In the unpolarized case, for  $e^+ +e^- \rightarrow h+ \bar q(\mathrm{jet})+X$,
there are only two twist-3 azimuthal asymmetries, $\langle \cos\varphi\rangle_U$ and $\langle \sin\varphi\rangle_U$,
and two twist-4 azimuthal asymmetries $\langle \cos2\varphi\rangle_U$ and $\langle \sin2\varphi\rangle_U$.
Two of them (the cosines) are parity conserved and the other two are parity violated.

For hadron polarization averaged over the azimuthal angle,
we have leading twist contributions with twist-4 addenda to the helicity $\langle\lambda_h\rangle$, the spin alignment $\langle S_{LL}\rangle$
and the transverse components with respect to the hadron-jet plane,
i.e., $\langle S_{T}^n\rangle$, $\langle S_{T}^t\rangle$, $\langle S_{LT}^n\rangle$,
$\langle S_{LT}^t\rangle$, $\langle S_{TT}^{nn}\rangle$ and $\langle S_{TT}^{nt}\rangle$.
For the transverse components with respect to the lepton-hadron plane,
i.e.,  $\langle S_{T}^x\rangle$, $\langle S_{T}^y\rangle$, $\langle S_{LT}^x\rangle$, $\langle S_{LT}^y\rangle$,
we have twist-3 contributions, while $\langle S_{TT}^{xx}\rangle$ and $\langle S_{TT}^{xy}\rangle$ only have twist-4 contributions.

The four-quark correltators also contribute at twist-4.
The contributions take the same modes as those at the leading twist
hence just addenda to the corresponding leading twist contributions.

For the inclusive reaction $e^+ +e^- \rightarrow h+X$, we have leading twist contributions with twist-4 addenda to
the helicity $\langle\lambda_h\rangle^{(in)}$, the spin alignment $\langle S_{LL}\rangle^{(in)}$.
For the transverse components with respect to the lepton-hadron plane,
we have twist-3 contributions to $\langle S_{T}^x\rangle^{(in)}$, $\langle S_{T}^y\rangle^{(in)}$, $\langle S_{LT}^x\rangle^{(in)}$, and $\langle S_{LT}^y\rangle^{(in)}$;
but only twist-4 contributions to $\langle S_{TT}^{xx}\rangle^{(in)}$ and $\langle S_{TT}^{xy}\rangle^{(in)}$.

The results are presented for $e^+e^-$-annihilation at the $Z$-pole where parity conserved and parity violated structure functions contribute.
These results reduce to those for $e^+e^-$-annihilation via virtual photon ($\gamma^*$) if we make the replacement of $c_V$ by $e_q$ and $c_A=0$.

We also suggest a method for a rough estimation of twist-4 contributions based on the leading twist fragmentation functions.
From the estimation, we see that the twist-4 contributions could be very significant and have large influences on extracting leading twist FFs from the data.

\section*{Acknowledgements}

We thank Shu-yi Wei and Yu-kun Song for helpful discussions.
This work was supported in part by the Major State Basic Research Development Program in China (No. 2014CB845406),
the National Natural Science Foundation of China (Nos. 11375104 and 11675092),
and the CAS Center for Excellence in Particle Physics (CCEPP).


\begin{thebibliography}{0}

\bibitem{Airapetian:1999tv}
  A.~Airapetian {\it et al.}  [HERMES Collaboration],
  ``Observation of a single spin azimuthal asymmetry in semiinclusive pion electro production,''
  Phys.\ Rev.\ Lett.\  {\bf 84}, 4047 (2000) [hep-ex/9910062].

\bibitem{Airapetian:2004tw}
  A.~Airapetian {\it et al.} [HERMES Collaboration],
  ``Single-spin asymmetries in semi-inclusive deep-inelastic scattering on a transversely polarized hydrogen target,''
  Phys.\ Rev.\ Lett.\  {\bf 94}, 012002 (2005)
  doi:10.1103/PhysRevLett.94.012002
  [hep-ex/0408013].

\bibitem{Airapetian:2009ae}
  A.~Airapetian {\it et al.}  [HERMES Collaboration],
  ``Observation of the Naive-T-odd Sivers Effect in Deep-Inelastic Scattering,''
  Phys.\ Rev.\ Lett.\  {\bf 103}, 152002 (2009),
 [arXiv:0906.3918 [hep-ex]].

\bibitem{Airapetian:2010ds}
  A.~Airapetian {\it et al.} [HERMES Collaboration],
  ``Effects of transversity in deep-inelastic scattering by polarized protons,''
  Phys.\ Lett.\ B {\bf 693}, 11 (2010)
  doi:10.1016/j.physletb.2010.08.012
  [arXiv:1006.4221 [hep-ex]].

\bibitem{Airapetian:2012yg}
  A.~Airapetian {\it et al.} [HERMES Collaboration],
  ``Azimuthal distributions of charged hadrons, pions, and kaons produced in deep-inelastic scattering off unpolarized protons and deuterons,''
  Phys.\ Rev.\ D {\bf 87}, no. 1, 012010 (2013)
  doi:10.1103/PhysRevD.87.012010
  [arXiv:1204.4161 [hep-ex]].

\bibitem{Alexakhin:2005iw}
  V.~Y.~Alexakhin {\it et al.} [COMPASS Collaboration],
  ``First measurement of the transverse spin asymmetries of the deuteron in semi-inclusive deep inelastic scattering,''
  Phys.\ Rev.\ Lett.\  {\bf 94}, 202002 (2005)
  doi:10.1103/PhysRevLett.94.202002
  [hep-ex/0503002].

\bibitem{Ageev:2006da}
  E.~S.~Ageev {\it et al.}  [COMPASS Collaboration],
  ``A New measurement of the Collins and Sivers asymmetries on a transversely polarised deuteron target,''
  Nucl.\ Phys.\ B {\bf 765}, 31 (2007)
  [hep-ex/0610068].

\bibitem{Alekseev:2008aa}
  M.~Alekseev {\it et al.} [COMPASS Collaboration],
  ``Collins and Sivers asymmetries for pions and kaons in muon-deuteron DIS,''
  Phys.\ Lett.\ B {\bf 673}, 127 (2009)
  doi:10.1016/j.physletb.2009.01.060
  [arXiv:0802.2160 [hep-ex]].

\bibitem{Alekseev:2010rw}
  M.~G.~Alekseev {\it et al.} [COMPASS Collaboration],
  ``Measurement of the Collins and Sivers asymmetries on transversely polarised protons,''
  Phys.\ Lett.\ B {\bf 692}, 240 (2010)
  doi:10.1016/j.physletb.2010.08.001
  [arXiv:1005.5609 [hep-ex]].


\bibitem{Adolph:2012sn}
  C.~Adolph {\it et al.} [COMPASS Collaboration],
  ``Experimental investigation of transverse spin asymmetries in muon-p SIDIS processes: Collins asymmetries,''
  Phys.\ Lett.\ B {\bf 717}, 376 (2012)
  doi:10.1016/j.physletb.2012.09.055
  [arXiv:1205.5121 [hep-ex]].

\bibitem{Adolph:2012sp}
  C.~Adolph {\it et al.} [COMPASS Collaboration],
  ``II ¨C Experimental investigation of transverse spin asymmetries in ¦Ì -p SIDIS processes: Sivers asymmetries,''
  Phys.\ Lett.\ B {\bf 717}, 383 (2012)
  doi:10.1016/j.physletb.2012.09.056
  [arXiv:1205.5122 [hep-ex]].

\bibitem{Adolph:2014pwc}
  C.~Adolph {\it et al.} [COMPASS Collaboration],
  ``Measurement of azimuthal hadron asymmetries in semi-inclusive deep inelastic scattering off unpolarised nucleons,''
  Nucl.\ Phys.\ B {\bf 886}, 1046 (2014)
  doi:10.1016/j.nuclphysb.2014.07.019
  [arXiv:1401.6284 [hep-ex]].

\bibitem{Adolph:2014zba}
  C.~Adolph {\it et al.} [COMPASS Collaboration],
  ``Collins and Sivers asymmetries in muon production of pions and kaons off transversely polarised protons,''
  Phys.\ Lett.\ B {\bf 744}, 250 (2015)
  doi:10.1016/j.physletb.2015.03.056
  [arXiv:1408.4405 [hep-ex]].


\bibitem{Avakian:2010ae}
  H.~Avakian {\it et al.}  [CLAS Collaboration],
  ``Measurement of Single and Double Spin Asymmetries in Deep Inelastic Pion Electroproduction with a Longitudinally Polarized Target,''
  Phys.\ Rev.\ Lett.\  {\bf 105}, 262002 (2010),
  [arXiv:1003.4549 [hep-ex]].

\bibitem{Aghasyan:2011ha}
  M.~Aghasyan {\it et al.},
  ``Precise measurements of beam spin asymmetries in semi-inclusive $\pi^0$ production,''
  Phys.\ Lett.\ B {\bf 704}, 397 (2011)
  doi:10.1016/j.physletb.2011.09.044
  [arXiv:1106.2293 [hep-ex]].

\bibitem{Qian:2011py}
  X.~Qian {\it et al.}  [Jefferson Lab Hall A Collaboration],
  ``Single Spin Asymmetries in Charged Pion Production from Semi-Inclusive Deep Inelastic Scattering on a Transversely Polarized $^3$He Target,''
  Phys.\ Rev.\ Lett.\  {\bf 107}, 072003 (2011)
  [arXiv:1106.0363 [nucl-ex]].

\bibitem{Huang:2011bc}
  J.~Huang {\it et al.}  [Jefferson Lab Hall A Collaboration],
  ``Beam-Target Double Spin Asymmetry $A_{LT}$ in Charged Pion Production from Deep Inelastic Scattering on a Transversely Polarized He-3 Target at 1.4
  Phys.\ Rev.\ Lett.\  {\bf 108}, 052001 (2012)
  [arXiv:1108.0489 [nucl-ex]].

\bibitem{Zhang:2013dow}
  Y.~Zhang {\it et al.} [Jefferson Lab Hall A Collaboration],
  ``Measurement of pretzelosity asymmetry of charged pion production in Semi-Inclusive Deep Inelastic Scattering on a polarized $^3$He target,''
  Phys.\ Rev.\ C {\bf 90}, no. 5, 055209 (2014)
  doi:10.1103/PhysRevC.90.055209
  [arXiv:1312.3047 [nucl-ex]].


\bibitem{Zhao:2014qvx}
  Y.~X.~Zhao {\it et al.} [Jefferson Lab Hall A Collaboration],
  ``Single spin asymmetries in charged kaon production from semi-inclusive deep inelastic scattering on a transversely polarized $^3He$ target,''
  Phys.\ Rev.\ C {\bf 90}, no. 5, 055201 (2014)
  doi:10.1103/PhysRevC.90.055201
  [arXiv:1404.7204 [nucl-ex]].



\bibitem{Abe:2005zx}
  K.~Abe {\it et al.} [Belle Collaboration],
  ``Measurement of azimuthal asymmetries in inclusive production of hadron pairs in $e^+ e^-$ annihilation at Belle,''
  Phys.\ Rev.\ Lett.\  {\bf 96}, 232002 (2006)
  doi:10.1103/PhysRevLett.96.232002
  [hep-ex/0507063].

\bibitem{Seidl:2008xc}
  R.~Seidl {\it et al.} [Belle Collaboration],
  ``Measurement of Azimuthal Asymmetries in Inclusive Production of Hadron Pairs in $e^+e^-$ Annihilation at $\sqrt{s} = 10.58$ GeV,''
  Phys.\ Rev.\ D {\bf 78}, 032011 (2008)
  Erratum: [Phys.\ Rev.\ D {\bf 86}, 039905 (2012)]
  doi:10.1103/PhysRevD.78.032011, 10.1103/PhysRevD.86.039905
  [arXiv:0805.2975 [hep-ex]].

\bibitem{Vossen:2011fk}
  A.~Vossen {\it et al.} [Belle Collaboration],
  ``Observation of transverse polarization asymmetries of charged pion pairs in $e^+e^-$ annihilation near $\sqrt{s}=10.58$ GeV,''
  Phys.\ Rev.\ Lett.\  {\bf 107}, 072004 (2011)
  doi:10.1103/PhysRevLett.107.072004
  [arXiv:1104.2425 [hep-ex]].


\bibitem{TheBABAR:2013yha}
  J.~P.~Lees {\it et al.} [BaBar Collaboration],
  ``Measurement of Collins asymmetries in inclusive production of charged pion pairs in $e^+e^-$ annihilation at BABAR,''
  Phys.\ Rev.\ D {\bf 90}, no. 5, 052003 (2014)
  doi:10.1103/PhysRevD.90.052003
  [arXiv:1309.5278 [hep-ex]].

\bibitem{Ablikim:2015pta}
  M.~Ablikim {\it et al.} [BESIII Collaboration],
  ``Measurement of azimuthal asymmetries in inclusive charged dipion production in $e^+e^-$ annihilations at $\sqrt{s}$ = 3.65 GeV,''
  Phys.\ Rev.\ Lett.\  {\bf 116}, no. 4, 042001 (2016)
  doi:10.1103/PhysRevLett.116.042001
  [arXiv:1507.06824 [hep-ex]].




\bibitem{Barone:2010zz}
  For a recent review see e.g., V.~Barone, F.~Bradamante and A.~Martin,
  ``Transverse-spin and transverse-momentum effects in high-energy processes,''
  Prog.\ Part.\ Nucl.\ Phys.\  {\bf 65}, 267 (2010),
  doi:10.1016/j.ppnp.2010.07.003
  [arXiv:1011.0909 [hep-ph]].


\bibitem{Liang:2015nia}  For a recent review, see e.g.,
  Z.~t.~Liang,
  ``Three Dimensional Imaging of the Nucleon --- TMD (Theory and Phenomenology),''
  Int.\ J.\ Mod.\ Phys.\ Conf.\ Ser.\  {\bf 40}, 1660008 (2016)
  doi:10.1142/S2010194516600089  [arXiv:1502.03896 [hep-ph]];
 also
  K.~b.~Chen, S.~y.~Wei and Z.~t.~Liang,
  ``Three Dimensional Imaging of the Nucleon and Semi-Inclusive High Energy Reactions,''
  Front.\ Phys.\ (Beijing) {\bf 10}, no. 6, 101204 (2015)
  doi:10.1007/s11467-015-0477-x  [arXiv:1506.07302 [hep-ph]].


\bibitem{Mulders:1995dh}
  P.~J.~Mulders and R.~D.~Tangerman,
  ``The Complete tree level result up to order $1/Q$ for polarized deep inelastic leptoproduction,''
  Nucl.\ Phys.\ B {\bf 461}, 197 (1996)
  Erratum: [Nucl.\ Phys.\ B {\bf 484}, 538 (1997)]
  doi:10.1016/S0550-3213(96)00648-7, 10.1016/0550-3213(95)00632-X
  [hep-ph/9510301].

\bibitem{Boer:1997nt}
  D.~Boer and P.~J.~Mulders,
  ``Time reversal odd distribution functions in leptoproduction,''
  Phys.\ Rev.\ D {\bf 57}, 5780 (1998)
  doi:10.1103/PhysRevD.57.5780
  [hep-ph/9711485].

\bibitem{Bacchetta:2004zf}
  A.~Bacchetta, P.~J.~Mulders and F.~Pijlman,
  ``New observables in longitudinal single-spin asymmetries in semi-inclusive DIS,''
  Phys.\ Lett.\ B {\bf 595}, 309 (2004)
  doi:10.1016/j.physletb.2004.06.052
  [hep-ph/0405154].

\bibitem{Bacchetta:2006tn}
  A.~Bacchetta, M.~Diehl, K.~Goeke, A.~Metz, P.~J.~Mulders and M.~Schlegel,
  ``Semi-inclusive deep inelastic scattering at small transverse momentum,''
  JHEP {\bf 0702}, 093 (2007)
  doi:10.1088/1126-6708/2007/02/093
  [hep-ph/0611265].



\bibitem{Liang:2006wp}
  Z.~t.~Liang and X.~N.~Wang,
  ``Azimuthal and single spin asymmetry in deep-inelastic lepton-nucleon scattering,''
  Phys.\ Rev.\ D {\bf 75}, 094002 (2007)
  [hep-ph/0609225].

\bibitem{Song:2010pf}
 Y.~k.~Song, J.~h.~Gao, Z.~t.~Liang and X.~N.~Wang,
  ``Twist-4 contributions to the azimuthal asymmetry in SIDIS,''
  Phys.\ Rev.\ D {\bf 83}, 054010 (2011)
  doi:10.1103/PhysRevD.83.054010
  [arXiv:1012.4179 [hep-ph]].

\bibitem{Song:2013sja}
  Y.~k.~Song, J.~h.~Gao, Z.~t.~Liang and X.~N.~Wang,
  ``Azimuthal asymmetries in semi-inclusive DIS with polarized beam and/or target and their nuclear dependences,''
  Phys.\ Rev.\ D {\bf 89}, no. 1, 014005 (2014)
  doi:10.1103/PhysRevD.89.014005
  [arXiv:1308.1159 [hep-ph]].

\bibitem{Yang:2016qsf}
  Y.~Yang and Z.~Lu,
  ``Polarized $\Lambda$ hyperon production in semi-inclusive deep inelastic scattering off an unpolarized nucleon target,''
  Phys.\ Rev.\ D {\bf 95}, no. 7, 074026 (2017)
  doi:10.1103/PhysRevD.95.074026
  [arXiv:1611.07755 [hep-ph]].

\bibitem{Boer:1997mf}
  D.~Boer, R.~Jakob and P.~J.~Mulders,
  ``Asymmetries in polarized hadron production in $e^+ e^-$ annihilation up to order $1/Q$,''
  Nucl.\ Phys.\ B {\bf 504}, 345 (1997)
  doi:10.1016/S0550-3213(97)00456-2
  [hep-ph/9702281].

\bibitem{Boer:2008fr}
  D.~Boer,
  ``Angular dependences in inclusive two-hadron production at BELLE,''
  Nucl.\ Phys.\ B {\bf 806}, 23 (2009)
  doi:10.1016/j.nuclphysb.2008.06.011
  [arXiv:0804.2408 [hep-ph]].

\bibitem{Wei:2013csa}
  S.~y.~Wei, Y.~k.~Song and Z.~t.~Liang,
  ``Higher twist contribution to fragmentation function in inclusive hadron production in $e^+e^-$ annihilation,''
  Phys.\ Rev.\ D {\bf 89}, no. 1, 014024 (2014)
  doi:10.1103/PhysRevD.89.014024
  [arXiv:1309.4191 [hep-ph]].

\bibitem{Wei:2014pma}
  S.~Y.~Wei, K.~b.~Chen, Y.~k.~Song and Z.~t.~Liang,
  ``Leading and higher twist contributions in semi-inclusive $e^{+}e^{-}$ annihilation at high energies,''
  Phys.\ Rev.\ D {\bf 91}, no. 3, 034015 (2015)
  doi:10.1103/PhysRevD.91.034015
  [arXiv:1410.4314 [hep-ph]].

\bibitem{Chen:2016moq}
  K.~b.~Chen, W.~h.~Yang, S.~y.~Wei and Z.~t.~Liang,
  ``Tensor polarization dependent FFs and $e^+e^-\rightarrow V \pi X$ at high energies,''
  Phys.\ Rev.\ D {\bf 94}, no. 3, 034003 (2016)
  doi:10.1103/PhysRevD.94.034003
  [arXiv:1605.07790 [hep-ph]].

\bibitem{Pitonyak:2013dsu}
  D.~Pitonyak, M.~Schlegel and A.~Metz,
  ``Polarized hadron pair production from electron-positron annihilation,''
  Phys.\ Rev.\ D {\bf 89}, no. 5, 054032 (2014)
  doi:10.1103/PhysRevD.89.054032
  [arXiv:1310.6240 [hep-ph]].


\bibitem{Wei:2016far}
  S.~y.~Wei, Y.~k.~Song, K.~b.~Chen and Z.~t.~Liang,
  ``Twist-4 contributions to semi-inclusive deeply inelastic scatterings with polarized beam and target,''
  Phys.\ Rev.\ D {\bf 95}, no. 7, 074017 (2017)
  doi:10.1103/PhysRevD.95.074017
  [arXiv:1611.08688 [hep-ph]].

\bibitem{Ellis:1982wd}
  R.~K.~Ellis, W.~Furmanski and R.~Petronzio,
  ``Power Corrections to the Parton Model in QCD,''
  Nucl.\ Phys.\ B {\bf 207}, 1 (1982).

\bibitem{Ellis:1982cd}
 R.~K.~Ellis, W.~Furmanski and R.~Petronzio,
  ``Unraveling Higher Twists,''
  Nucl.\ Phys.\ B {\bf 212}, 29 (1983).

\bibitem{Qiu:1990xxa}
  J.~-w.~Qiu and G.~F.~Sterman,
  ``Power corrections in hadronic scattering. 1. Leading $1/Q^2$ corrections to the Drell-Yan cross-section,''
  Nucl.\ Phys.\ B {\bf 353}, 105 (1991).

\bibitem{Qiu:1990xy}
 J.~-w.~Qiu and G.~F.~Sterman,
  ``Power corrections to hadronic scattering. 2. Factorization,''
  Nucl.\ Phys.\ B {\bf 353}, 137 (1991).

\bibitem{footnote}
We note that the convention used in defining FFs in this paper is the same as that for defining PDFs in \cite{Wei:2016far} 
but slightly different from that for FFs in \cite{Chen:2016moq}.
The differences are: 
3 twist-2 ($G_{1L}$, $D_{1T}^\perp$, $G_{1TT}^\perp$) and 7 twist-3 FFs ($G^\perp$, $D_L^\perp$, $D_T^\perp$,
$G_L^\perp$, $G_{LL}^\perp$, $D_{LT}^\perp$, $D_{TT}^{\prime\perp}$) have opposite sign;
the 6 twist-3 transverse polarization FFs ($D_T$, $G_T$, $D_{LT}$, $G_{LT}$, $D_{TT}^{\perp}$, $G_{TT}^{\perp}$)
in \cite{Chen:2016moq} correspond to the combination such as $\frac{k_\perp^2}{2M^2}D_T^\perp - D_T$.


\bibitem{deFlorian:1997zj}
  D.~de Florian, M.~Stratmann and W.~Vogelsang,
  ``QCD analysis of unpolarized and polarized Lambda baryon production in leading and next-to-leading order,''
  Phys.\ Rev.\ D {\bf 57}, 5811 (1998)
  doi:10.1103/PhysRevD.57.5811
  [hep-ph/9711387].

\bibitem{Albino:2008fy}
 S.~Albino, B.~A.~Kniehl and G.~Kramer,
  ``AKK Update: Improvements from New Theoretical Input and Experimental Data,''
 Nucl.\ Phys.\ B {\bf 803}, 42 (2008)
  doi:10.1016/j.nuclphysb.2008.05.017
  [arXiv:0803.2768 [hep-ph]].


\bibitem{Chen:2016iey}
  K.~b.~Chen, W.~h.~Yang, Y.~j.~Zhou and Z.~t.~Liang,
  ``Energy dependence of hadron polarization in $e^+e^-\to hX$ at high energies,''
  Phys.\ Rev.\ D {\bf 95}, no. 3, 034009 (2017)
  doi:10.1103/PhysRevD.95.034009
  [arXiv:1609.07001 [hep-ph]].


\bibitem{Qiu:1988dn}
  J.~W.~Qiu,
  ``Twist Four Contributions to the Parton Structure Functions,''
  Phys.\ Rev.\ D {\bf 42}, 30 (1990).
  doi:10.1103/PhysRevD.42.30

\end{thebibliography}
 \end{document}